\documentclass[a4paper,10pt]{article}
\usepackage[utf8]{inputenc}
\usepackage{authblk}

\usepackage{graphicx}
\usepackage{epstopdf}
\usepackage{booktabs}
\usepackage{todonotes}
\usepackage{lineno,hyperref}
\usepackage{upgreek}
\usepackage{units}
\usepackage{multirow}
\usepackage{array}
\usepackage{amssymb}	% Mathematische Symbole
\usepackage{amsmath}	% Mathematische Symbole
\usepackage{textcomp}
\usepackage{pdfpages}

\usepackage{caption}
\usepackage{subcaption}

\usepackage{cleveref}	% better references

\title{Mapping the material distribution of a complex structure in an electron beam}

\author[1,2]{Luise Poley}
\author[3]{Ulf Stolzenberg}
\author[3]{Benjamin Schwenker}
\author[3]{Ariane Frey}
\author[4]{Peter Göttlicher}
\author[5]{Carlos Marinas}
\author[4]{Marcel Stanitzki}
\author[1,2]{Bernd Stelzer}

\affil[1]{Department of Physics, Simon Fraser University, University Drive, Burnaby, Canada}
\affil[2]{TRIUMF, Wesbrook Mall, Vancouver, Canada}
\affil[3]{II. Physikalisches Institut, Georg-August-Universität Göttingen, Friedrich-Hund-Platz, Göttingen, Germany}
\affil[4]{Deutsches Elektronen-Synchrotron, Notkestraße, Hamburg, Germany}
\affil[5]{Instituto de F\'{\i}sica Corpuscular, CSIC-Universidad de Valencia, c/ Catedr\'{a}tico Jos\'{e} Beltr\'{a}n, Paterna, Spain}

\begin{document}

\maketitle

\begin{abstract}
The simulation and analysis of High Energy Physics experiments require a realistic simulation of the detector material and its distribution. The challenge is to describe all active and passive parts of large scale detectors like ATLAS in terms of their size, position and material composition. The common method for estimating the radiation length by weighing individual components, adding up their contributions and averaging the resulting material distribution over extended structures provides a good general estimate, but can deviate significantly from the material actually present.

A method has been developed to assess its material distribution with high spatial resolution using the reconstructed scattering angles and hit positions of high energy electron tracks traversing an object under investigation. The study presented here shows measurements for an extended structure with a highly inhomogeneous material distribution. The structure under investigation is an End-of-Substructure-card prototype designed for the ATLAS Inner Tracker strip tracker - a PCB populated with components of a large range of material budgets and sizes.

The measurements presented here summarise requirements for data samples and reconstructed electron tracks for reliable image reconstruction of large scale, inhomogeneous samples, choices of pixel sizes compared to the size of features under investigation as well as a bremsstrahlung correction for high material densities and thicknesses. 
\end{abstract}

\tableofcontents

\section{Introduction}  

Physics requirements on the vertex and momentum resolution of particles set limits on the mass of vertex detectors for the next generation of collider experiments~\cite{Cooper}. Localised interactions of particles in the detector material perturb the particle momentum, produce hadronic vertices with new secondary particles or lead to photon conversions~\cite{Aaboud:2017pjd, Sirunyan_2018}. To meet the conflicting goals of minimising detector mass and also incorporating cooling, power distribution and readout electronics, an integrated system design approach is needed~\cite{demarteau}. 
 
Multiple scattering leads to random deflections of charged particles traversing the detector. The magnitude of multiple scattering depends on the local material budget measured in units of the material constant $X_{0}$. In order to study the material composition of objects, a method has been developed to estimate high-resolution 2D images of the present material $X/X_{0}$ using high resolution tracking telescopes in test beam facilities~\cite{Stolzenberg2019}. The amount of material can be extracted from the multiple scattering deflections by using an appropriate theoretical model. The good spatial resolution of the method opens a way to measure the overall scattering effect in small areas. It can consequently be used to map fully passive materials very precisely and has previously been used for small areas of low mass pixel and strip detectors for the Belle~II vertex detector~\cite{vci_2017_x0}. 

For a large-scale application of the developed methods, a prototype End-of-Sub\-struc\-ture (EoS) card~\cite{EOS}, an electronics board developed for the new Inner Tracker~\cite{ITk} of the ATLAS detector~\cite{ATLAS}, was chosen. The EoS card is a comparably large structure (about \unit[12$\times$5]{cm}) located within the ATLAS tracking volume, which was designed to accommodate the combined ATLAS ITk requirements of high reliability, radiation hardness and minimum material. However, the functionalities of the EoS card (described in more detail in section~\ref{sec:theEOS}) require the use of several components containing high-density materials. For a reliable description of the material distribution in the detector simulation, a measurement of the material budget of the EoS card (results shown in section~\ref{sec:res}) was therefore necessary.

\section{Multiple scattering models}
\label{sec:multiple_scattering}

Multiple scattering is an unavoidable source of uncertainty in tracking and vertex detector systems. The thickness $X$ of the material divided by a material dependent quantity $X_0$, the radiation length constant, determines the magnitude of scattering angles which can be expected. The radiation length constant is approximately given by \cite{pdg_1998} 

\begin{align}
X_0&=\frac{716.4\,\mathrm{g/}\mathrm{cm}^2 \, A}
{Z  \left( Z + 1\right) \, \ln{\left(\nicefrac{287}{\sqrt{Z}}\right)}} \label{eq:X0}
\end{align}

\noindent The $X$/$X_0$ measurements presented here are based on the reconstruction of multiple scattering angles on a target material. Multiple scattering is the net effect of a large number of individual Coulomb single scatterings on nuclei of a traversed material and causes a direction change of the particle trajectory which can be expressed by two projected angles $\vartheta_u$ and $\vartheta_v$. There are several models, such as the Moliere~\cite{Moliere_multiple_scattering} and Highland~\cite{Highland_1975} model, which can be used to describe the multiple scattering angle distribution.\\
\newline
\noindent
Moliere introduced a fully analytical model that describes the Gaussian core as well as the non Gaussian tails of the multiple scattering distribution~\cite{Moliere_multiple_scattering}. The Highland model describes the Gaussian core of the multiple scattering angle distribution. The standard deviation, called the Highland width $\sigma_\mathrm{HL}$ is given by \cite{pdg_2018,Lynch_1991}:

\begin{align}
\sigma_\mathrm{HL}&=\frac{0.0136\,\mathrm{GeV}}{\beta p} \, \sqrt{\frac{X}{X_0}} \left( 1 + 0.038\, \ln\left( \frac{X}{X_0} \right) \right) \label{eq:highlandwidth}
\end{align}

\section{Radiation length measurements}
\label{sec:radiationlength_measurements}

\begin{figure}[htp]
           \centering
       \includegraphics[width=0.79\linewidth]{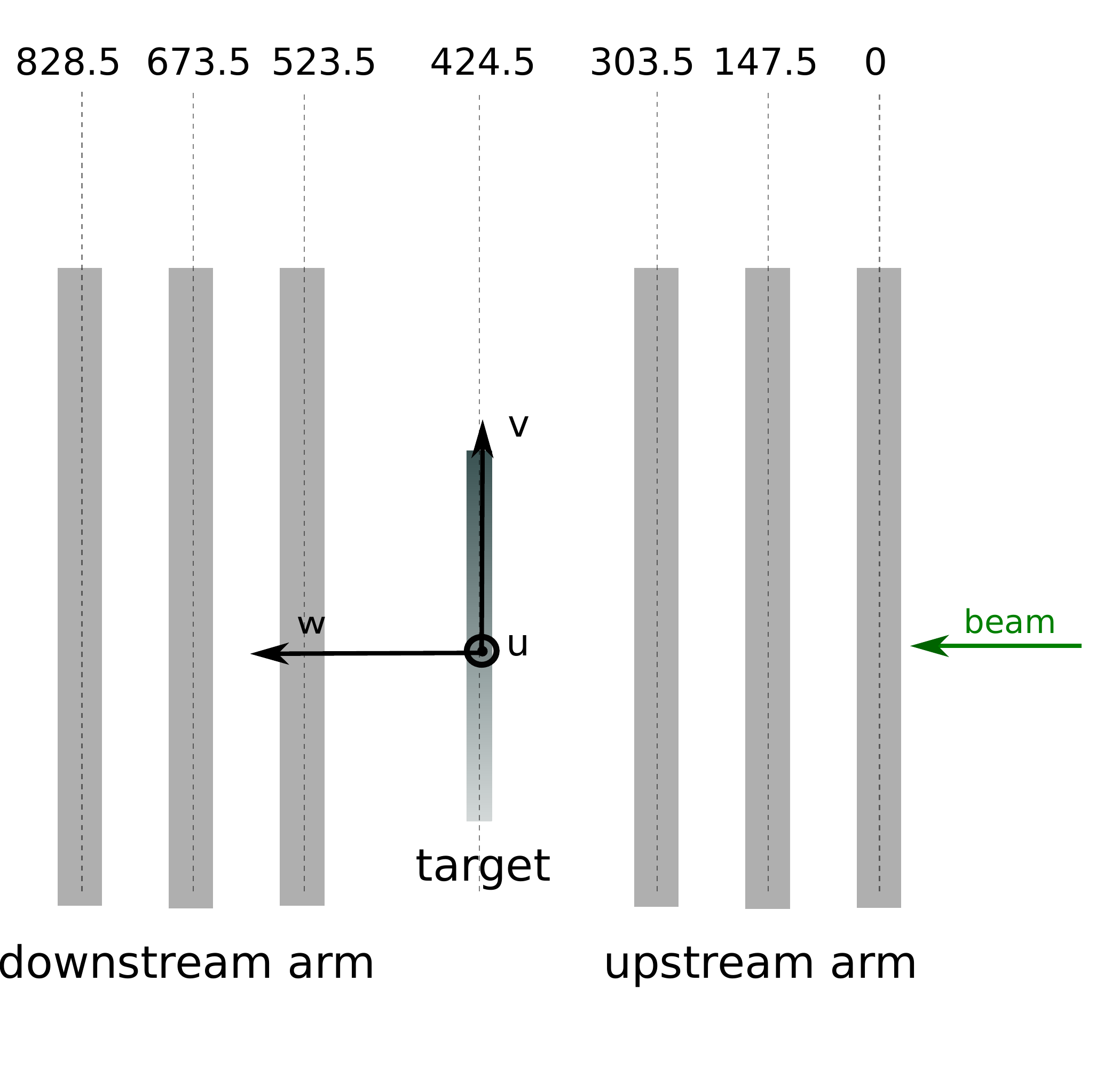}
       \caption{{\small Schematic drawing of the telescope setup. The scattering target, the EoS card, is placed between two telescope arms which consist of arrays of three position sensitive M26 sensors each.}}
       \label{fig:telescope_setup}
\end{figure}

The radiation length measurement approach presented here has already been used to estimate material profiles of modules of the vertex detector of Belle II~\cite{vci_2017_x0}. More details on the approach can be found in \cite{Stolzenberg2019}. The results presented here are based on measurements conducted at the DESY II testbeam facility~\cite{DESYII} with a \unit[2]{GeV} electron beam. The data analysis was conducted with the TestBeam SoftWare (TBSW) framework~\cite{tbsw}. An EUDET like reference telescope ~\cite{Rubinskiy2012923}, consisting of six Mimosa26 (M26) sensors, was used to measure the particle trajectories. A schematic drawing of a generic telescope setup is depicted in figure \ref{fig:telescope_setup}.\\

\noindent
The basic idea of the radiation length estimation is to reconstruct multiple scattering angles of charged particle transitions on a target plane centered in a high-resolution telescope. Details about data analysis steps such as clustering, tracking and alignment can be found in \cite{Schwenker14,Stolzenberg2019}. The particle hits on the sensors of the upstream telescope arm are used in a forward Kalman filter to determine \textit{in-states} $\eta_\mathrm{in}$ and their covariance matrix $C_\mathrm{in}$ of the particle tracks on the target plane. The forward filter is constructed such that the multiple scattering in air between telescope planes is taken into account by adding virtual scattering planes halfway between the telescope planes. The material of a telescope plane\footnote{The overall thickness of each plane consists of \unit[50]{$\mu$m} silicon and \unit[25]{$\mu$m} carbon} is treated in an asymmetric way and contributes to the extrapolation to the next downstream plane. In other words, the \textit{in-state} at the scattering target is possible without making an assumption on the material budget at that plane. Hits in the downstream telescope arm are used in a time-reversed Kalman filter pass starting from the last telescope plane. The unknown target material does therefore not enter in the prediction of the track-states $\eta_\mathrm{in}$ and $\eta_\mathrm{out}$ or their covariance matrices.\\
\newline
\noindent Without a magnetic field the tracks can be expressed by straight lines with kinks on every material surface. The track states on the target plane with a Cartesian coordinate system ($u$,$v$,$w$)\footnote{$u$ and $v$ are arranged in the target plane, while $w$ is perpendicular to the surface} can be therefore expressed by two intersection coordinates $u$ and $v$, two slopes $m_u=\nicefrac{\mathrm{d}u}{\mathrm{d}w}$ and $m_v=\nicefrac{\mathrm{d}v}{\mathrm{d}w}$ and the particle charge $q$ divided by the particle momentum $p$, for example:

\begin{align}
	\eta&= \left(\begin{array}{ccccc} m_u &	                                                                m_v &
	                                              u &
	                                              v & 
	                                              q/p
	                             \end{array}\right) 
	\label{eq:intrackstate}
\end{align}
\noindent
In- and out-states are matched according to a distance criterion. Calculating the difference between the slope components of $\eta_\mathrm{out}$ and $\eta_\mathrm{in}$ yields two projected scattering angles $\vartheta_u$ and $\vartheta_v$. The corresponding projected angle uncertainties can be determined from error propagation when using the appropriate covariance matrix entries. These uncertainties correspond to the telescope angle resolution $\sigma_\mathrm{reso}$, which is an important parameter during the calibration procedure. In order to perform spatially resolved measurements each reconstructed angle must be associated with a intersection position. This position is calculated as a weighted\footnote{Using the inverse intersection variances as weights} mean of the intersection entries in the track states.\\
\newline
\noindent
The target plane is divided into small image pixels and the scattering angles are filled into per pixel histograms using the estimated track intersection coordinates. The reconstructed scattering angle distribution of the pixel centered on position ($u$,$v$) can be described by a model function consisting of convolution between a multiple scattering distribution $f^{uv}_\mathrm{msc}$ and $f_\mathrm{err}$ for the angle reconstruction error.  
\begin{align}
f^{uv}_\mathrm{reco}\left(\vartheta|\nicefrac{X}{X_0}, p\right)=f^{uv}_\mathrm{msc}\left(\vartheta|\nicefrac{X}{X_0}, p\right)
\otimes f_\mathrm{err}\left(\vartheta\right) \quad \mathrm{,} \label{eq:fitmodel} 
\end{align}
\noindent
The scattering angle distribution $f^{uv}_\mathrm{msc}$ depends on the local radiation length fraction $X$/$X_0$ of the traversed material and the beam momentum. The density $f^{uv}_\mathrm{msc}$ can be expressed either by the Highland model (\cref{eq:highlandwidth}) or the Moliere model \cite{Moliere_multiple_scattering}. The angular resolution function of the telescope $f_\mathrm{err}$ is approximated by a normal distribution with zero mean and width $\sigma_{err}$. To first order, the angle resolution is calculated from the  uncertainties of the projected scattering angles for in and out states. More precisely, it can be measured from data during calibration as the width of the reconstructed angle distribution when no material is inserted between the telescope arms. The  calibration factor $\lambda$ is the ratio between the calculated angle resolution and the calibrated one.  

In the Highland theory, we have an explicit formula for the fit model

\begin{align}
f^{uv}_\mathrm{reco}&=\sum_{i} w_{i,u,v} \mathcal{N}(\mu_{i,u,v}, \sigma_{i,u,v}) \quad \mathrm{,} \label{eq:fitmodel_hl} 
\end{align}

where $ \mathcal{N}(0, 1)$ is a standard normal distribution. The sigma for component $i$ in pixel $(u,v)$ is parametrised as 

\begin{align}
\sigma_{i,u,v}= \sigma_{HL}(p(u,v), X/X_{0,i,u,v})^{2} +  \lambda^{2}  \sigma_\mathrm{err}^{2}    \quad \mathrm{.} \label{eq:sigma_hl_fit} 
\end{align}

A parametrisation for the beam momentum $p$ is provided later in this section. The weights $w_i$ can be interpreted as the fractional areas of a material with thickness $X/X_{0,i}$ covering part of the pixel area. For homogeneous pixels, a single component suffices to describe the central part of the reconstructed angle distribution. For non homogeneous pixel, a mixture with two components can improve the results. The total material length per pixel is found as a weighed mean of the individual components. 
\newline
\noindent
As the radiation length estimation depends on accurate measurements of small scattering angles \footnote{$\mathcal{O}(100\,$\textmu{}rad$)$ for thin materials such as 100$\,$\textmu{}m silicon}, the position of the telescope planes and the intrinsic resolution of the M26 sensors must be known precisely. To achieve this requirement a telescope calibration step, which includes telescope alignment, measurements of the cluster resolution and hot pixel masking must be conducted. Details on the telescope calibration can be found in \cite{Schwenker14,Stolzenberg2019}. Additionally, in order to correct for small uncertainties in the telescope alignment, cluster calibration and assumption of the nominal beam energy and telescope length a calibration measurement with a well known material profile must be conducted.\\
\newline
In the frame of this calibration several measurement regions with a known radiation length $X$/$X_0$ are defined. The reconstructed angle distributions in these measurement regions can be described by an individual model function in \cref{eq:fitmodel}. For each of these function the radiation length parameter $X$/$X_0$ is fixed to the corresponding value. Calibration parameters, such as $\lambda$, $\kappa$ and the exact linear dependence of the beam energy $p(u,v)$ are shared by the model functions of all measurement regions and can be determined by performing a simultaneous fit of the reconstructed multiple scattering distributions.\\

\noindent 

The beam energy used in the fit model (see eq.~\ref{eq:fitmodel}) allows linear momentum gradients and a scale correction $\kappa$ from the nominal momentum $p_{\mathrm{nom}}$ set by the test beam facility. 

\begin{align}
p\left(u,v\right)&=\kappa \left( p_{\mathrm{nom}} + \nabla_u\,p \cdot u + \nabla_v\,p \cdot v \right) \quad \, \mathrm{.} \label{eq:linear_beamenergy} 
\end{align}

In addition to the beam energy gradients, energy losses of electrons due to bremsstrahlung in the scattering target decreases the momentum depending in the local material distribution. The average energy loss can be modeled by the Bethe Heitler model~\cite{BetheHeitler} and depends only on the local radiation length $X$/$X_0$. Guided by experimental results, we used the following parametrisation to correct for the effect of bremsstrahlung in our method.

\begin{align}
p_\mathrm{wm}^\epsilon(\nicefrac{X}{X_0},u,v)&=\left(1-\epsilon\right)\, p\left(u,v\right)+\epsilon\, p\left(u,v\right)\,\exp{\left(\nicefrac{-X}{X_0}(u,v)\right)} \label{eq:momentum_weightedmean} 
\end{align}

 The beam momentum $p_\mathrm{wm}$ is a weighed mean of the momentum before and after traversing the scattering target. The weight parameter $\epsilon$ is a measure for the impact of the overall bremsstrahlung energy loss on the multiple scattering distribution and has to be determined from measurements on high $X$/$X_0$ materials. However, it is not a calibration parameter and the task of determining $\epsilon$ is separate from the radiation length calibration. Example measurements on aluminium and copper wedges with large radiation lengths values will be presented in section \ref{sec:teststructures}.

\begin{figure}[htp]
\centering
	\includegraphics[width=0.99\linewidth]{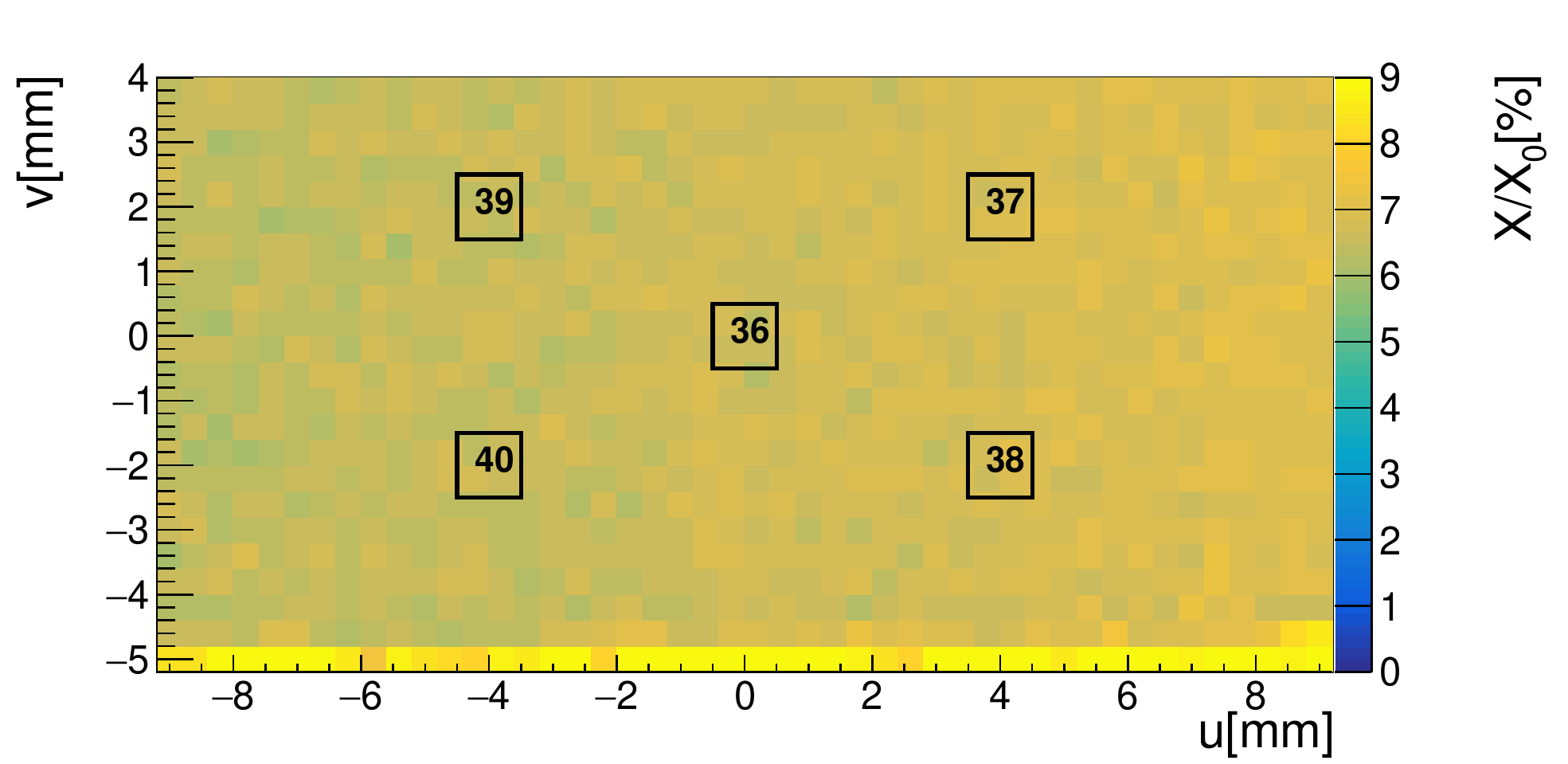}
	\caption{{\small Radiation length image of the {\unit[6]{mm}} aluminium target. The measurement areas providing scattering angle distributions for the simultaneous fit to determine the calibration factors are shown as black rectangles. The measurement areas are located in the central part of the image, where constant radiation length values are observed.}}
	\label{fig:calibration_measurementareas}
\end{figure}

For the data set presented here a calibration measurement was conducted with a set of aluminium targets with a constant thickness. The seven aluminium layers had thicknesses ranging from \unit[0.5 to 6]{mm} of aluminium. Additionally, scattering data without a scattering target were recorded. For each material five million six hit tracks were reconstructed. The number of scattering angles per measurement area was limited to 100 000 angles to ensure an identical statistical weight of each angle distribution during the simultaneous fit. The measurement with an air target is especially sensitive on the calibration parameter $\lambda$ which is connected to the telescope angle resolution. The calculated telescope angle resolution was found to be $\sigma_\mathrm{reso}=\,$179$\,$\textmu{}rad. In order to determine the linear dependence of the beam energy on the position on the target plane, measurement areas should cover the whole range of the beam spot. Here measurement areas at the four corners of the beam spot and a single measurement area near the center of the beam spot are used. Figure \ref{fig:calibration_measurementareas} depicts one of the calibration targets, aluminium with a thickness of 6$\,$mm, with the corresponding measurement areas marked by black and white rectangles.\\

\noindent The calibration parameters determined for $\epsilon$ values of 0.24, 0.30 and 0.36 are listed in \cref{table:calibration_parameters}. According to these calibration parameters, the calibrated telescope angle resolution $\lambda\,\sigma_\mathrm{reso}\approx181\pm1\,$\textmu{}rad deviates from the calculated telescope angle resolution $\sigma_\mathrm{reso}$ on the target plane by approximately 1$\,\%$. $\kappa$ was found to be approximately 0.99, which indicates that there was a small error in the assumed beam energy or telescope length. The beam energy gradient in $v$ direction is compatible with 0$\,\nicefrac{\mathrm{MeV}}{\mathrm{mm}}$, while the gradient in $u$ direction is $\nabla_u\,p=-10\pm1\nicefrac{\mathrm{MeV}}{\mathrm{mm}}$.

\begin{table}[htp]
	\centering
	\small
	\begin{tabular}{ccccc} \toprule
	 	$\epsilon$ & $\lambda$ & $\kappa$ & $\Delta_u p$ [MeV/mm]  & $\Delta_v p$ [MeV/mm]\\ \midrule
		0.24 & 1.011$\pm$0.005 & 0.988$\pm$0.003 & -9$\pm$2 & -1$\pm$3 \\
		 0.30  & 1.011$\pm$0.005 & 0.987$\pm$0.003 & -10$\pm$2 & -1$\pm$3 \\
		 0.36  & 1.012$\pm$0.005 & 0.985$\pm$0.003 & -10$\pm$2 & -1$\pm$3 \\ \bottomrule
		\hline
	\end{tabular}
	\caption{Calibration results for $\epsilon$ values of 0.24, 0.30 and 0.36. $\lambda$ is the calibration factor of the telescope resolution. $\kappa$ is the calibration factor of the scattering angle width and $\Delta_u p$ and $\Delta_v p$ correspond to beam energy gradients in $u$ and $v$ direction. The given errors correspond to the statistical uncertainties of the fit.\\}
	\label{table:calibration_parameters}
\end{table}

\begin{figure}[htp]
         		 \centering
	\begin{subfigure}[b]{0.490\textwidth}
		 \includegraphics[width=0.99\linewidth]{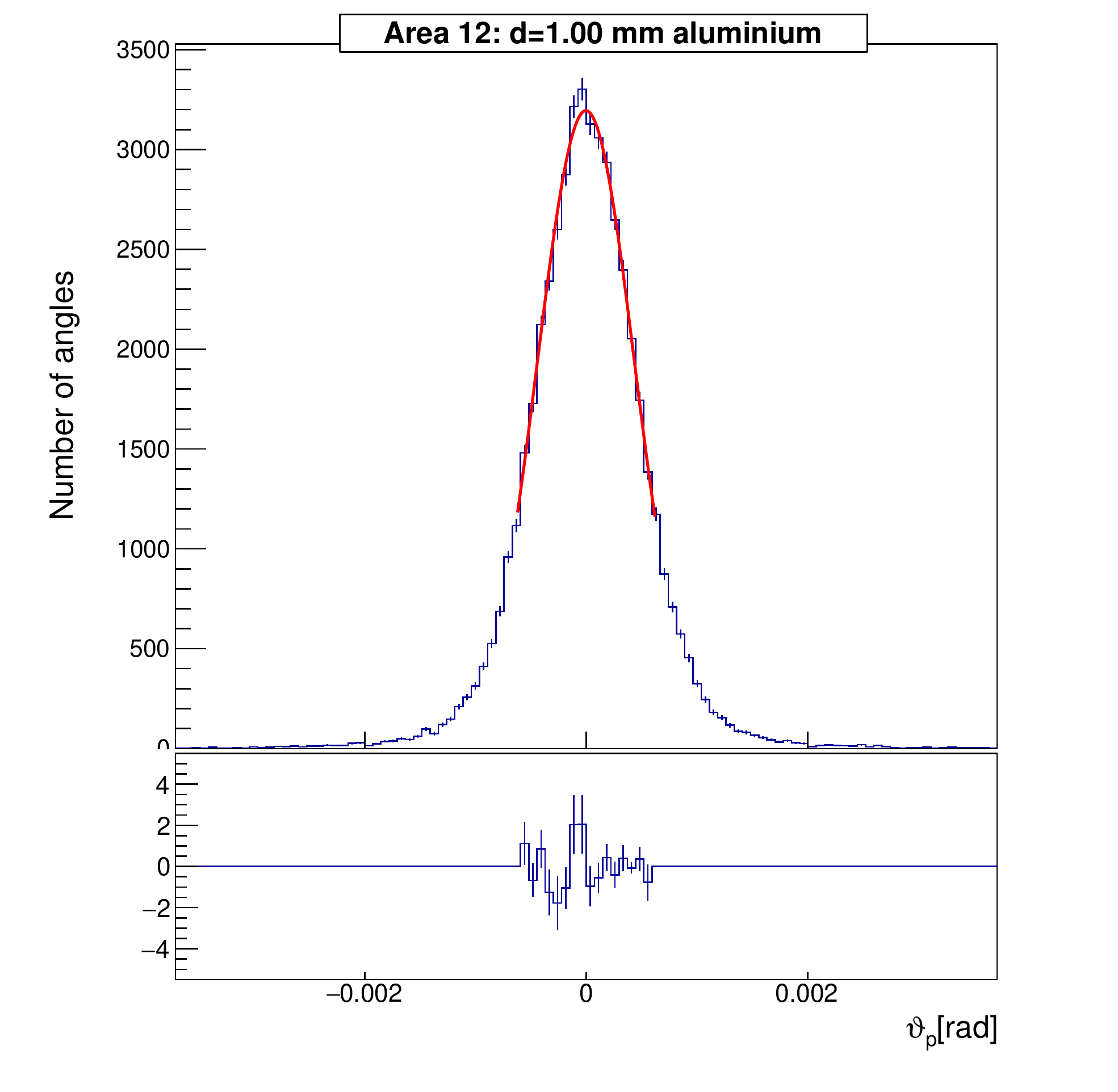}
  		 \label{fig:calibration_fits_area9}
  	\end{subfigure}
  	\begin{subfigure}[b]{0.490\textwidth}
		 \includegraphics[width=0.99\linewidth]{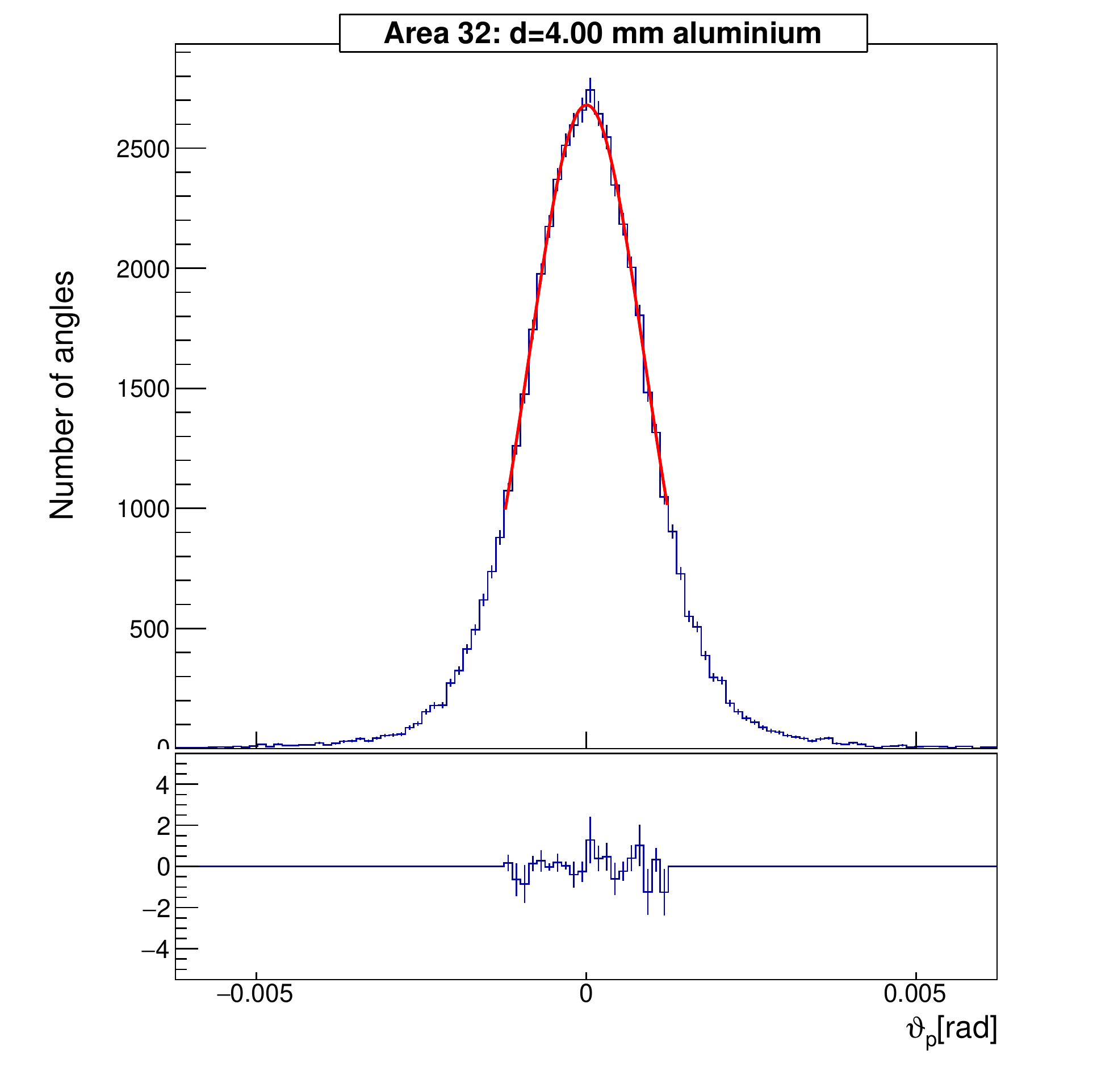}
  		 \label{fig:calibration_fits_area18}
  	\end{subfigure}
	\caption{{\small Global fit to obtain calibration constants $\kappa$, $\lambda$, $\Delta_u$ and $\Delta_v$. The fit minimises the joint $\chi^2$ for fitting the scattering angles in 30 measurement areas on aluminium plates with known thickness. The thickness is fixed and only the calibration constants are floated. The data and fit is is shown for two example areas and show good agreement. }}
	\label{fig:calibration_fits}
\end{figure}

Beam energy gradients in this direction have been observed frequently in our previous measurements~\cite{Stolzenberg2019}. The non-zero energy gradient in $u$ direction is caused by the beam energy selection process at DESY, which includes the horizontal spread of the multi-energetic particle beam via dipole magnet\cite{DESYII}. Two of the 40 scattering angle distributions used during the calibration process together with their fit functions are depicted in figure \ref{fig:calibration_fits}.

\begin{figure}[htp]
       \centering
       \includegraphics[width=0.99\textwidth,angle=270]{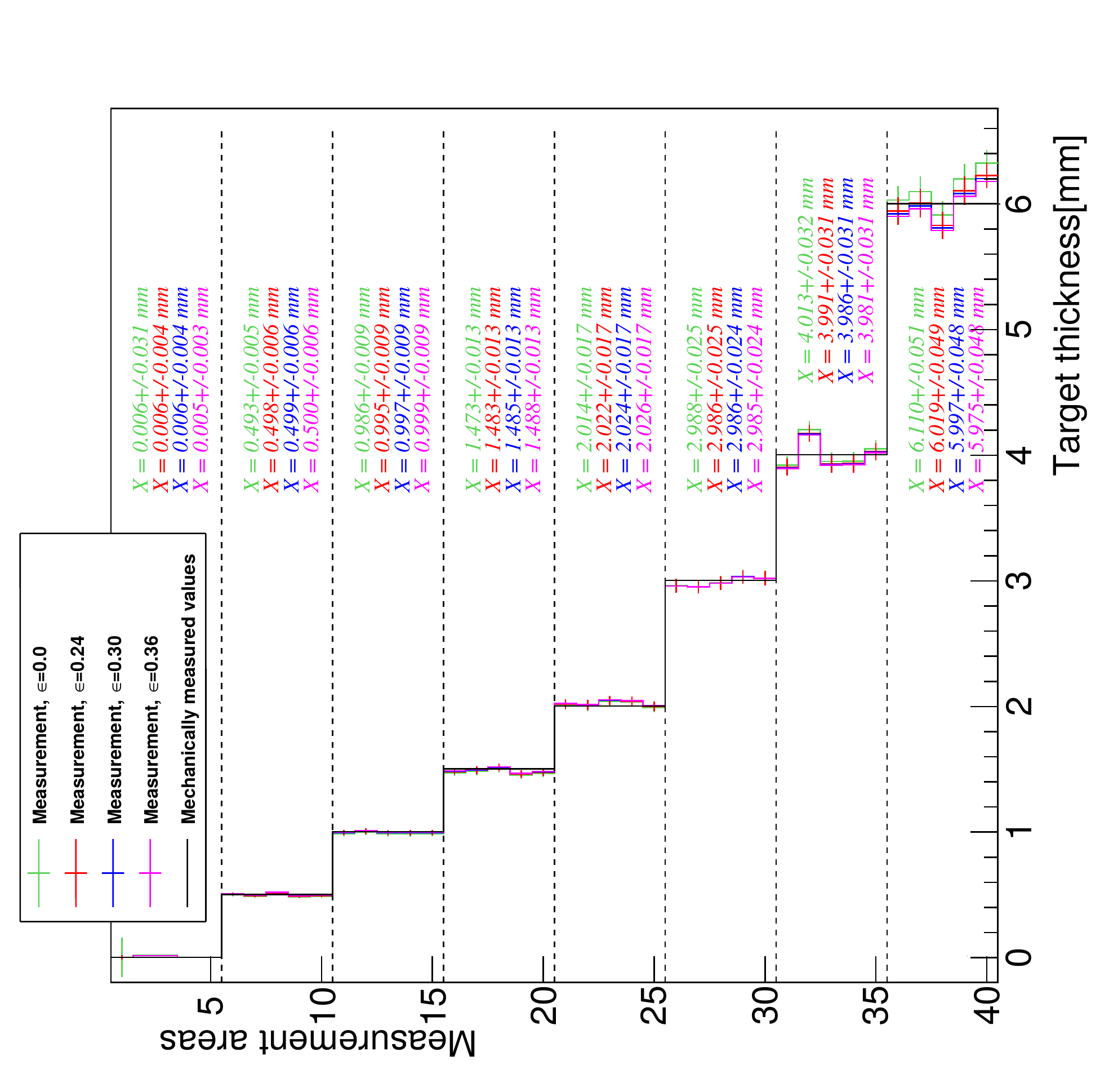}
       \caption{{\small Comparison between mechanically measured thickness values of the calibration targets with constant thicknesses (ranging from {\unit[0.5 to 6]{mm}} of aluminium) and the estimated thickness after the calibration for four different $\epsilon$ values. As can be seen the estimated thicknesses are compatible with the mechanically measured thicknesses within the limits of the statistical uncertainty for nearly all thicknesses. Only for a thickness of {\unit[6]{mm}} the $\epsilon=0$ estimated values show small deviations from the mechanical measurement.}}
       \label{fig:calibration_validationplot}
\end{figure}

\noindent
The presented $X$/$X_0$ calibration procedure contains a validation step. Using the determined calibration factors ($\lambda$, $\kappa$ and the beam energy gradients) the angle distributions, which were used during the calibration, are fitted to estimate the radiation length or analogously the aluminium thickness. The determined values should match the thickness assumptions used during the $X$/$X_0$ calibration. These estimated thicknesses together with the mechanically measured aluminium thickness values are depicted in figure \ref{fig:calibration_validationplot}. The results of the estimation show a good agreement with the mechanically measured values. As can be seen, the selected small differences in the $\epsilon$ values only have a minor effect on the thickness measurements. The deviations between measurements with different $\epsilon$ values are smaller than the statistical uncertainties. Only for a $\epsilon$ value of zero a small deviation at the largest aluminium thickness is observed. This is not surprising as the largest radiation length value employed during the calibration corresponds to \unit[6.7]{\%} and the bremsstrahlung correction only has a large impact for $X$/$X_0$ values beyond \unit[10]{\%}.

\section{Measurements with test structures}
\label{sec:teststructures}
\begin{figure}[htp]
         		 \centering
	\begin{subfigure}[b]{0.490\textwidth}
		 \includegraphics[width=0.99\linewidth]{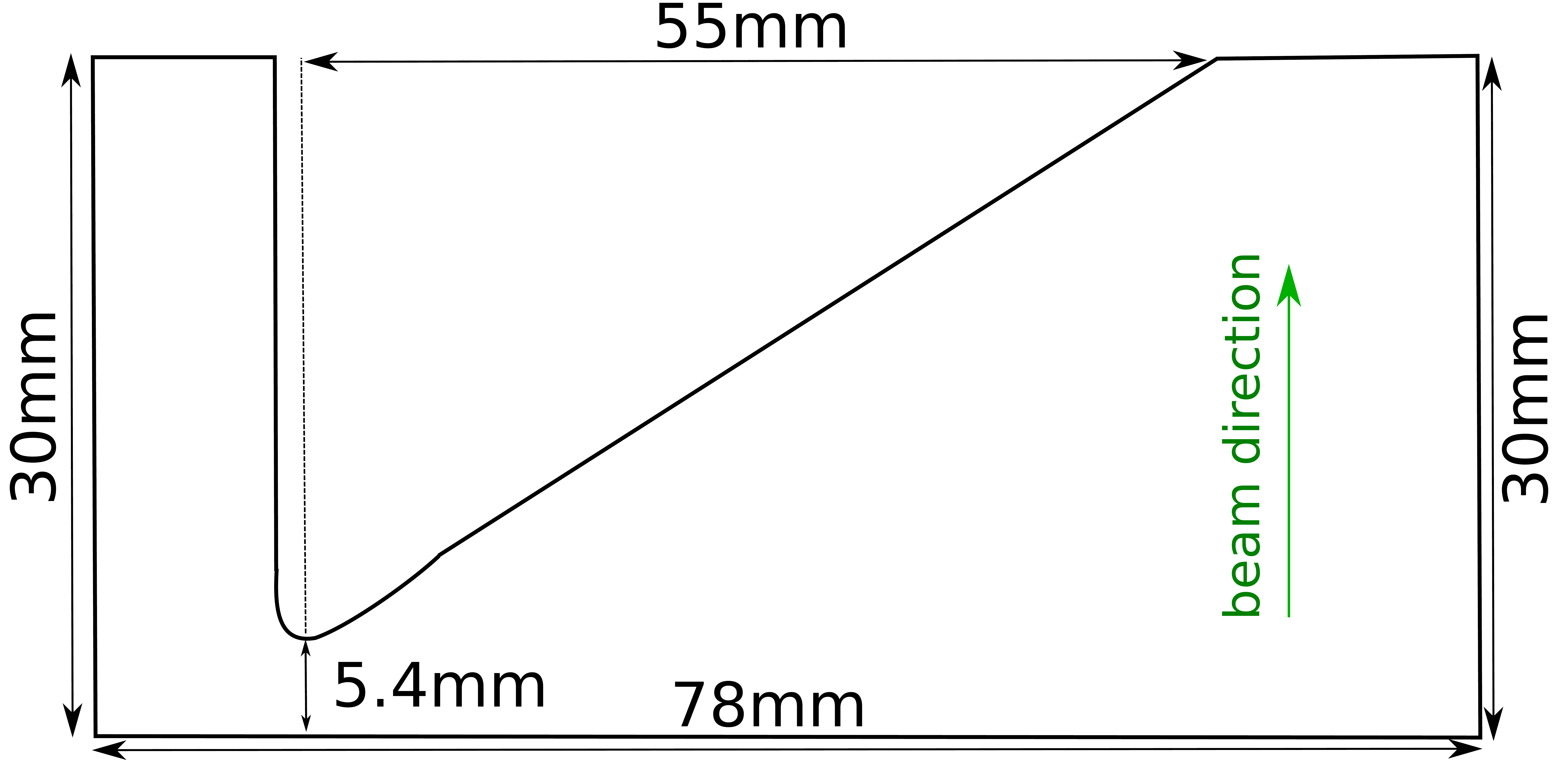}
  		 \label{fig:wedge_schematic}
  	\end{subfigure}
  	\begin{subfigure}[b]{0.490\textwidth}
         \includegraphics[width=0.99\linewidth]{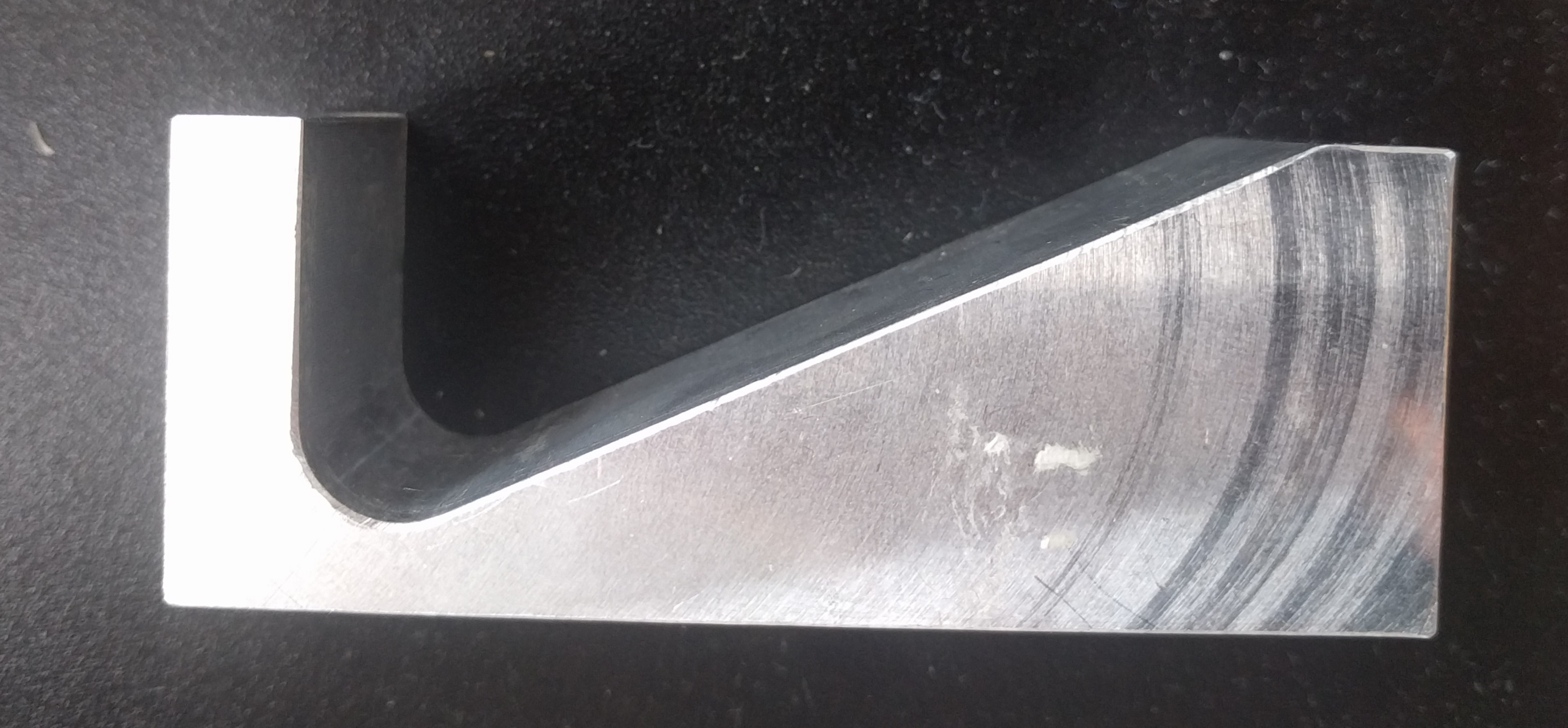}
  		 \label{fig:wedge_photo}
  	\end{subfigure}
	\caption{\small Schematic drawing and photograph of the measured aluminium wedge. The wedge has a continuous thickness profile ranging from approximately {\unit[5.4]{mm}} to 30$\,$mm of aluminium which corresponds to a $X$/$X_0$ value of roughly {\unit[34]{\%}}. Images from~\cite{Stolzenberg2019}}
	\label{fig:wedge}
\end{figure}

Test structures of aluminium and copper with a wedge like profile provide a test bench for the presented imaging algorithm. The samples are homogeneous and the thickness profile is accurately known. The analysis focuses on extracting quantitative radiation length profiles from scattering angles along the wedge and experimentally finding good values for the weight parameter $\epsilon$.The measurements where conducted in 2017 and 2018 at the DESY test beam facility.\\
\newline
\noindent A schematic drawing and a photograph of the wedge are depicted in figure \ref{fig:wedge}. The wedge ranges from \unit[6]{mm} at its thinnest point up to \unit[30]{mm} of aluminium, which corresponds to a radiation length $X$/$X_0$ of \unit[33.7]{\%}~\footnote{According to \cite{pdg_2018} the radiation length constant of aluminium is $X_0$ = \unit[88.97]{mm}}. The wedge is a rather large structure with a total length of \unit[78]{mm}. The slope angle of the wedge was determined to be $24.296^\circ$ with an optical coordinate measurement machine (optical CMM). Due to the limited size of the M26 sensitive area of approximately 10$\, \times \,$20 $\mathrm{mm}^2$ several measurements at different positions along the wedge had to be conducted. The distance between two measurement positions was \unit[7]{mm} and a total of eleven wedge positions were measured. Before the measurements calibration steps were conducted as described in the last section. 

\begin{figure}[htp]
\centering
    \includegraphics[width=0.99\linewidth]{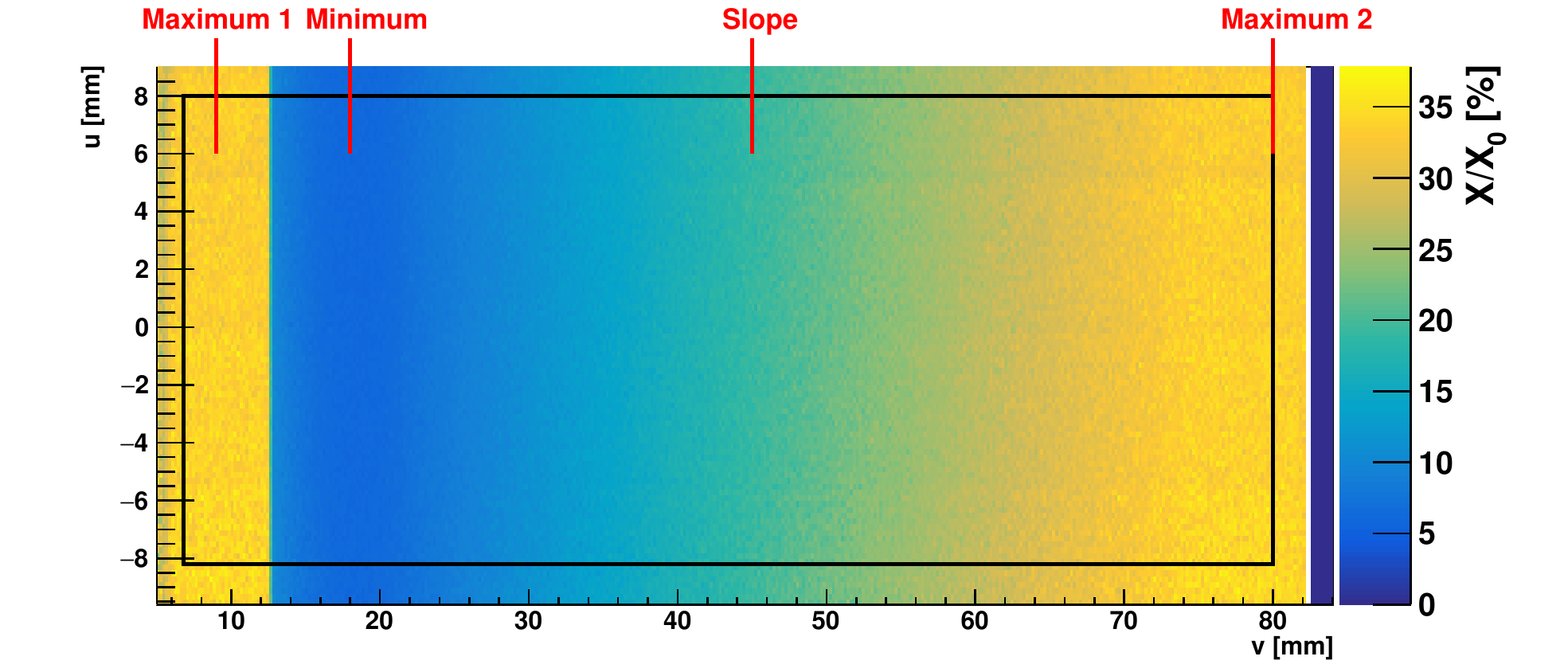}
	\caption{{\small Radiation length $X$/$X_0$ image of the aluminium wedge. Several features of the wedge and the image pixels used to generate the profile in figure \ref{fig:wedge_profiles} are indicated. Images from~\cite{Stolzenberg2019}}}
	\label{fig:wedge_image}
\end{figure}

\begin{figure}[htp]
\centering
    \includegraphics[width=0.99\linewidth]{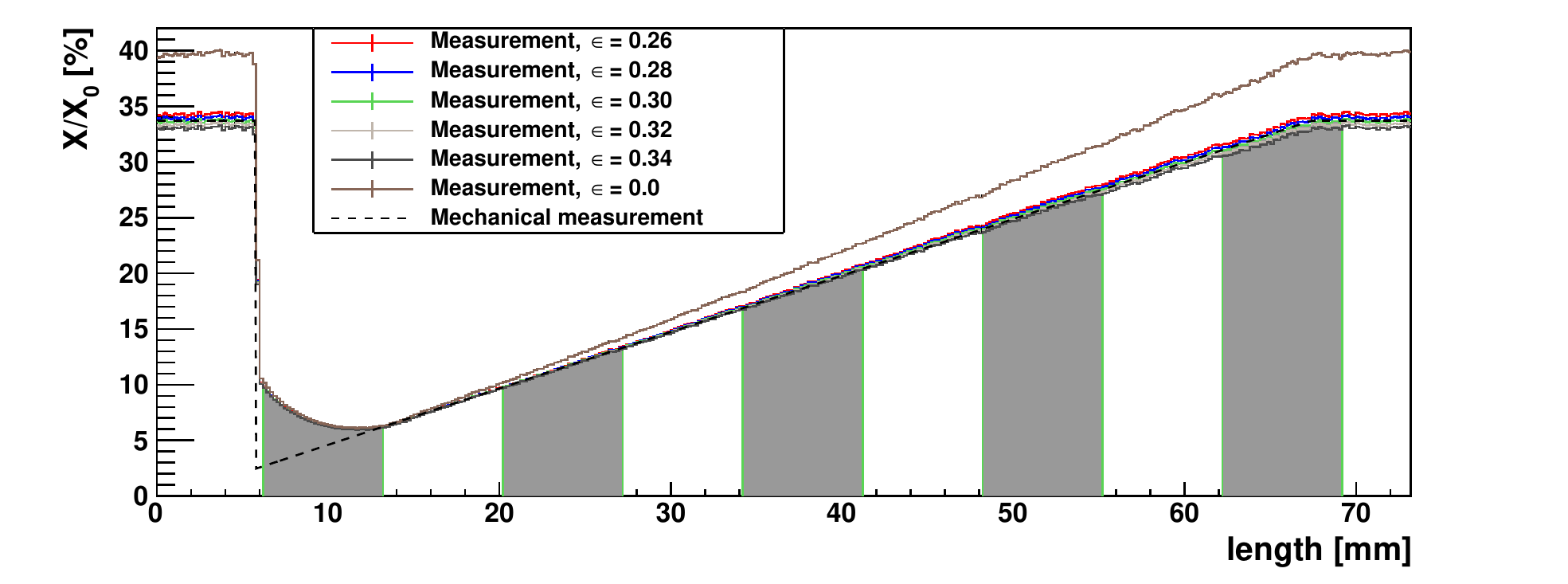}
	\caption{{\small Material profile of the aluminium wedge. The different measurement positions along the wedge are indicated by the alternating grey and white areas. The estimated $X$/$X_0$ values with epsilon 0.30 and 0.32 match the mechanically measured profile (indicated by the black dashed line) the best. The estimated $X$/$X_0$ values without the bremsstrahlung correction ($\epsilon=0$) show large deviations from the mechanical measurement, especially for large $X$/$X_0$ values. }}
	\label{fig:wedge_profiles}
\end{figure}

\noindent The $X$/$X_0$ image of the wedge is depicted in figure \ref{fig:wedge_image}. The image pixel inside the black rectangle are used to determine the radiation length along the slope of the wedge and compare them to values determined from the precision measurement of the slope angle. The material profiles for six different $\epsilon$ values are depicted in figure \ref{fig:wedge_profiles}. The alternating grey and white areas indicate the eleven different measurement regions along the wedge. As can be seen the $X$/$X_0$ estimation with epsilon values between 0.26 and 0.34 match the expected profile along the slope of the wedge quite well. The estimated radiation length values without bremsstrahlung correction, i.e. an epsilon of zero, show large deviations from the mechanically measured profile. \\

\begin{figure}[htp]
\centering
    \includegraphics[width=0.99\linewidth]{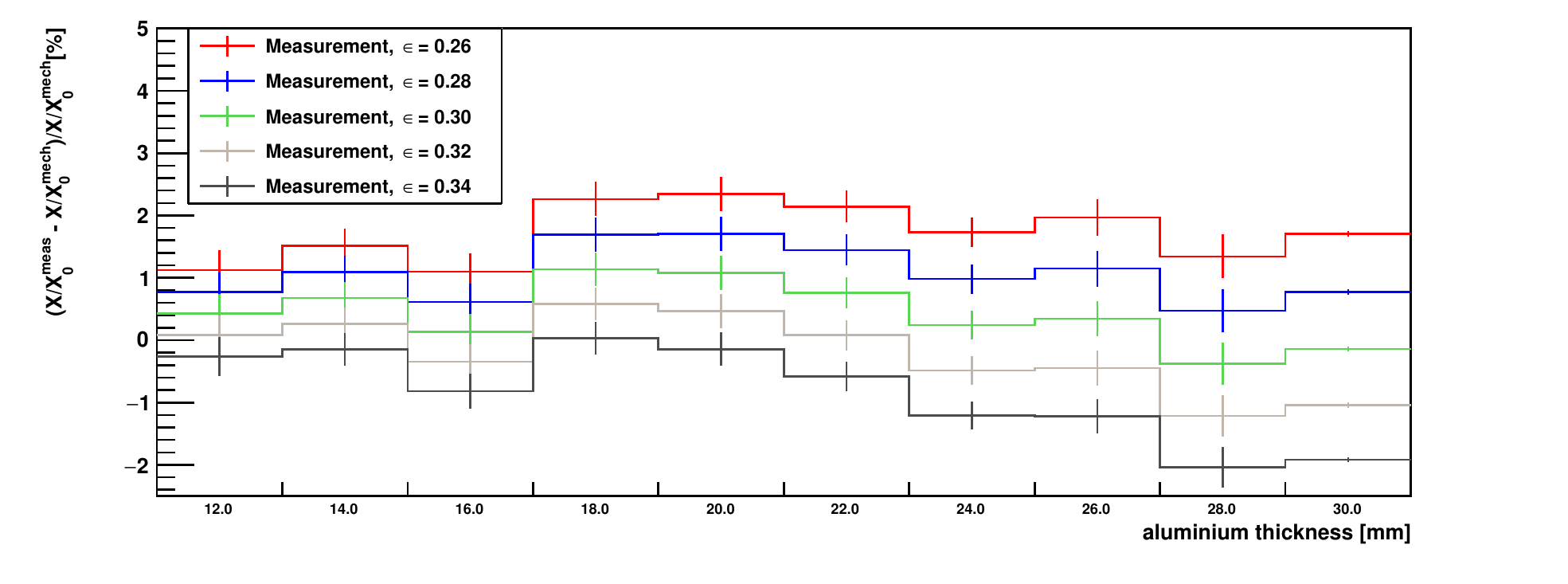}
	\caption{{\small Relative difference between the estimated radiation length values at specific points along the aluminium wedge and the corresponding values of the mechanical measurement. Estimated values with a weight parameter $\epsilon$ = 0.30 and 0.32 match the mechanically measured values the best. Accordingly, the optimal weight parameter lies between 0.30 and 0.32.}}
	\label{fig:epsilon_scan_difference}
\end{figure}

\noindent \Cref{fig:epsilon_scan_difference} displays the difference $\Delta \nicefrac{X}{X_0}=\left(\nicefrac{X}{X_0}\right)^\mathrm{meas}-\left(\nicefrac{X}{X_0}\right)^\mathrm{mech}$ between the estimated radiation length values $\left(\nicefrac{X}{X_0}\right)^\mathrm{meas}$ and the expected values based on mechanical measurements $\left(\nicefrac{X}{X_0}\right)^\mathrm{mech}$ as a function of the corresponding aluminium thickness of the wedge for five different $\epsilon$ values. As can be seen an $\epsilon$ value between 0.3 and 0.32 shows the best agreement with the expected values.\\

\noindent In order to confirm this weight factor and test its validity, a complementary measurement with a copper wedge was conducted~\cite{Stolzenberg2019}. The measurements confirmed our previous findings that for materials with large radiation length in the order of \unit{10}{\%} $X$/$X_0$, energy losses due to bremsstrahlung have to be included in radiation length estimation  to ensure unbiased $X$/$X_0$ measurements.

\section{Material distribution maps of extended structures}

\subsection{Structure under investigation}
\label{sec:theEOS}

The End-of-Substructure (EoS) card is a component designed for the Upgrade of the ATLAS Inner Detector: the ITk strip tracker~\cite{ITk}. The ITk strip tracker consists of four barrel layers in the central region and six end-cap discs in the forward region, each populated with silicon sensor modules mounted on both sides. Each barrel layer and end-cap disc will comprise identical substructures: staves in the central region, holding 14 modules on either side, and petals in the forward region, holding nine modules on either side.
EoS cards are mounted on both sides of each substructure to act as the interface between modules and off-detector electronics. For the foreseen detector geometry, particle tracks with a pseudo-rapidity $\eta$ between about 1.2 and 2.0, tracks are likely to traverse at least one EoS card.

The ITk is designed to have a material budget of less than \unit[2]{\% $X/X_{0}$}~\cite{ITk} along a particle track. However, locally, a higher material budget can not be avoided due to the need for certain components, such as copper coils and heat sinks on an EoS card. While locally high material budgets can not always be avoided, it is important to model their material budget correctly in tracking and tracking simulations.

\begin{figure}
    \centering
    \includegraphics[width=0.99\linewidth]{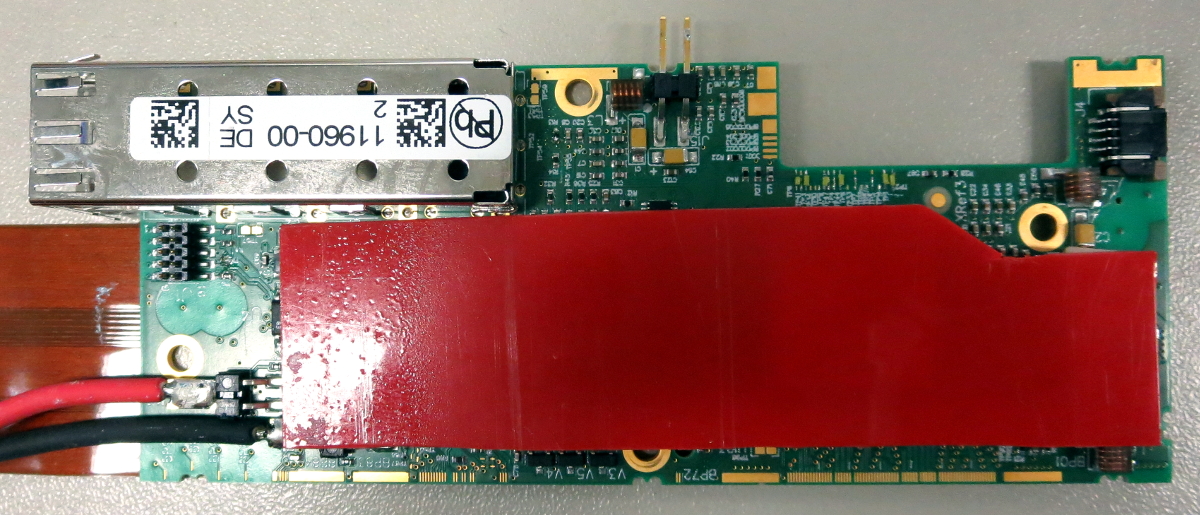}
    \caption{Photograph of the EoS card under investigation. In order to protect wire bonds on the chips mounted on the EoS, sensitive areas are protected by a cover (shown in red) on the circuit board.}
    \label{fig:EoS_foto}
\end{figure}

A prototype EoS card was therefore chosen as a test object for these material budget measurement studies: as an extended structure (about \unit[12$\times$5]{cm}, see figure~\ref{fig:EoS_foto}) with a large range of components it provides an opportunity to study material budget imaging for a range of structures and feature sizes. In addition, the determination of its overall material budget provides a valuable input for ITk tracker simulations.

The EoS card studied here is a populated printed circuit board (PCB). It is designed to distribute the control signals and power to the front-end modules and collects, serialises and transmits the up-stream data to an off-detector data acquisition system. For the prototype shown here, the active high data rate part is performed within an ASIC designed for the LHC experiments, the GBTX~\cite{GBTX}. Another ASIC, the GBT-SCA~\cite{GBT-SCA}, collects slowly changing environmental information. The communication to the off-detector electronics is performed with a bi-directional \unit[5]{Gbit/s} optical link. For the prototype, the electrical-optical conversion is done using commercial SFP$+$ plugins, which are planned to be replaced by low weight radiation hard optical modules for the final design. In addition to the active components, passive C-L-C pi-filters are used for different functional blocks.
 
\begin{figure}
  \centering
  \includegraphics[width=0.9\linewidth]{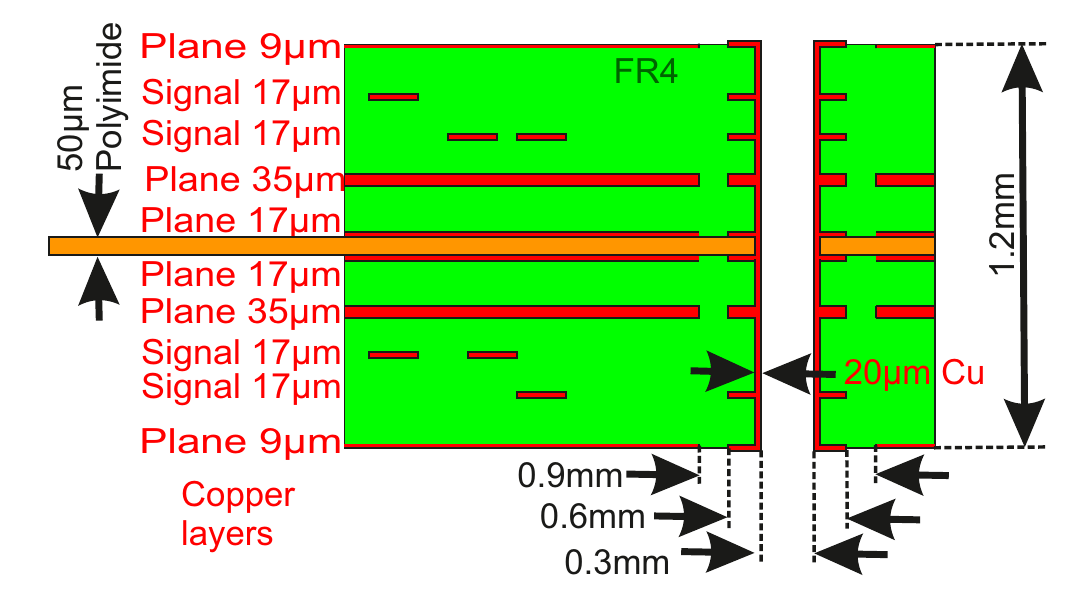}
  \caption{Layer stack of the EoS card and the structure of a through going interconnect (via).}
  \label{fig:EoS_Stack}
\end{figure}

The PCB itself is a multi-layer stack of interleaving glass-epoxy (FR4) and 10 copper foils (see figure ~\ref{fig:EoS_Stack}). It has a total thickness of \unit[1.2]{mm} (\unit[$\pm$10]{\%}). The central layer is a \unit[50]{$\upmu$m} polyimide layer. Most of the homogeneous material is made up by six copper layers - nearly full planes of individual thicknesses from \unit[9]{$\upmu$m} to \unit[35]{$\upmu$m}, adding up to a combined thickness of \unit[120]{$\upmu$m}.

In the transverse structure, the PCB layers are interrupted by vias, vertical connections between layers. Most of them are designed as vias connecting through the whole PCB (through vias). They have a central hole with a radius of \unit[0.15]{mm}, visible as low material areas of circular shape. Corresponding drills have a depth of \unit[1.2]{mm} and their cylindrical surfaces are plated with \unit[$\approx20$]{$\upmu$m} of copper, visible as areas with more material. An additional contribution to areas with visibly more material comes from copper rings with a radius of \unit[0.3]{mm} surrounding the drill holes in each of the ten layers (polyimide and copper), which add up to a total thickness of \unit[0.2]{mm} of copper. 

A different set of holes in the PCB corresponds to vias from the outer layer to layer 3, with a depth of only \unit[0.23]{mm} and the diameters of both copper rings and drill holes reduced by \unit[1]{mm} each.

Known features were used to compare the estimated material with the known material budget in the corresponding area (see section~\ref{subsec:known}).

\subsection{Measurement setup}

Measurements were conducted at Beamline 21 at the DESY II testbeam facility using electrons using the DATURA beam telescope~\cite{DATURA}. While other types of particle beams, e.g. muon beams, would be better suited for multiple scattering measurements due to lower energy losses and no bremsstrahlung, electron beams provide a good compromise between availability with high beam currents and traceable range of scattering angles.

Several data runs were performed for the mapping and data analysis of the structure under investigation. It is important to note that the measurement setup was unchanged between different measurements: a constant beam energy of \unit[2]{GeV} was used in combination of a telescope sensor threshold setting of 5 (corresponding to a threshold \unit[5]{$\sigma$} above the sensor noise). For all measurements related to the EoS card, the telescope planes were positioned as listed in table~\ref{tab:telplanes}.
    \begin{table}[htp]
    \centering
    \begin{tabular}{cc} \toprule
    Plane & Position $[$mm$]$ \\
    \midrule
    0 & 0\\
    1 & 147.5\\
    2 & 303.5\\
    3 & 523.5\\
    4 & 673.5\\
    5 & 828.5\\ \bottomrule
    \end{tabular}
    \caption{Positions of telescope planes (along the beam axis, with the first traversed plane 0 positioned at $z = 0$) for all material distribution measurements. The structure under investigation was placed between planes 2 and 3.}
    \label{tab:telplanes}
    \end{table}
    
Data runs with the following targets were performed (in accordance with the measurement routine explained in section~\ref{sec:radiationlength_measurements}):
\begin{itemize}
    \item no target ("air run") \\
    data taking without any target for telescope calibration
    \item calibration runs \\
    data taking with aluminium plates of known, precise thicknesses to calibrate calculated material using known samples, about 1 million telescope events per calibration target. The obtained reconstruction constants are summarised in table~\ref{tab:calconst}
    \item material distribution mapping \\
    Mapping the full EoS card by systematically moving the structures, mounted on precision translation stages, with respect to the beam. 5-6 million telescope events were collected per position.
    \item high statistics runs \\
    for two structures of particular interest (readout chips with a heat sink), large data samples were collected (about 50 million telescope events for a single stage position) in order to produce high-resolution maps
\end{itemize}
    \begin{table}[tp]
    \centering
    \begin{tabular}{cc} \toprule
    Constant & Value \\
    \midrule
    $\lambda$ & 1.044 \\
    $\kappa$ & 0.95 \\ %berechnet aus offset = 2.0993 -> 2 GeV/offset?
    $\Delta_u p$ & \unit[7.3]{MeV/mm} \\
    $\Delta_v p$ & \unit[0.8]{MeV/mm} \\ \bottomrule
    \end{tabular}
    \caption{Calibration constants calculated based on calibration measurements as defined in section~\ref{sec:radiationlength_measurements}. All measured constants are in good agreement with the values used to determine a reasonable bremsstrahlung correction.}
    \label{tab:calconst}
    \end{table}
    
\begin{figure}
    \centering
    \includegraphics[width=\linewidth]{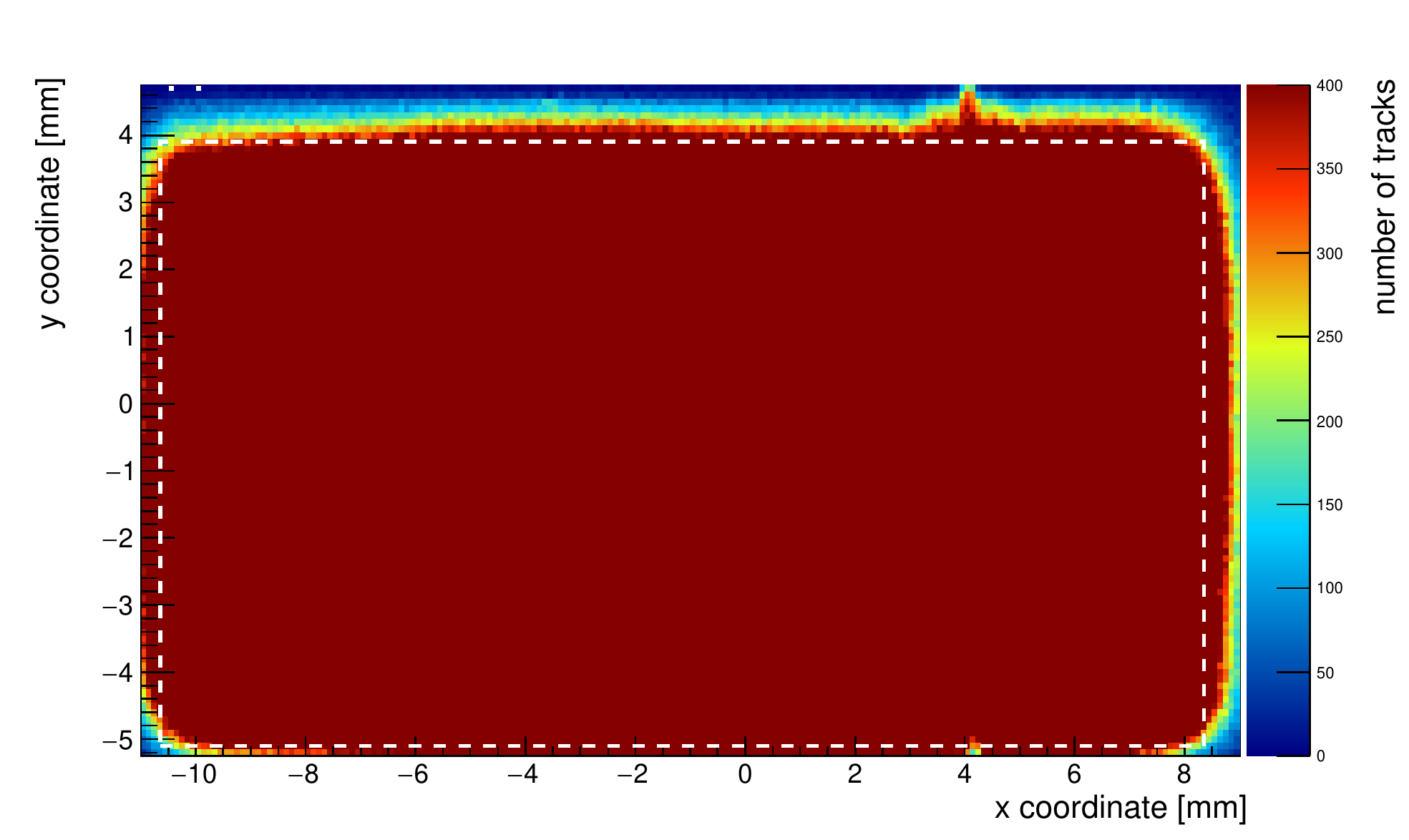}
    \caption{Number of reconstructed tracks per pixel in a material distribution map from 5 million events: 400 tracks per pixel were required for a reliable fit. The white frame shows the area assumed to be usable per stage position as a basis of estimate of the stepping size.}
    \label{fig:tracks}
\end{figure}

The size of an individual material distribution map is determined by the size of the beam telescope's MIMOSA-26 sensors~\cite{MIMOSA} ($\unit[1 \times 2]{\text{cm}^2}$~\cite{Rubinskiy2012923}), since only tracks passing through the active sensor area contribute to the reconstructed image (see figure~\ref{fig:tracks}). In order to use the full telescope sensor area, a beam collimator with the corresponding size ($\unit[1 \times 2]{\text{cm}^2}$) was chosen for these measurements.
In order to reconstruct the material budget using the fit described in section~\ref{sec:radiationlength_measurements}, a minimum number of 400 reconstructed tracks per pixel was required in order to consider the resulting result reliable. As a result, pixels at the edges of the imaging window are only reconstructed for large data samples, where sufficient tracks are found even in the edge regions of the image.
The stepping size between subsequent stage positions was therefore chosen to minimise overlap between adjacent images (and therefore minimise the number of stage positions where data had to be collected) while providing hermetic coverage between adjacent images.
Stage positions were therefore changed by \unit[19]{mm} in x and \unit[9]{mm} in y (see figure~\ref{fig:tracks}), with 5-6 million telescope events per stage position being collected. For each beam position, data was collected for \unit[45-55]{minutes} (for 4-5 million telescope events), at a trigger rate of about \unit[1.5]{kHz}. Mapping the full structure (excluding detail images mapped with larger data samples) required a total of \unit[39]{hours}.

\begin{figure}
\centering
 \begin{subfigure}{\textwidth}
  \centering
  \includegraphics[width=.9\linewidth]{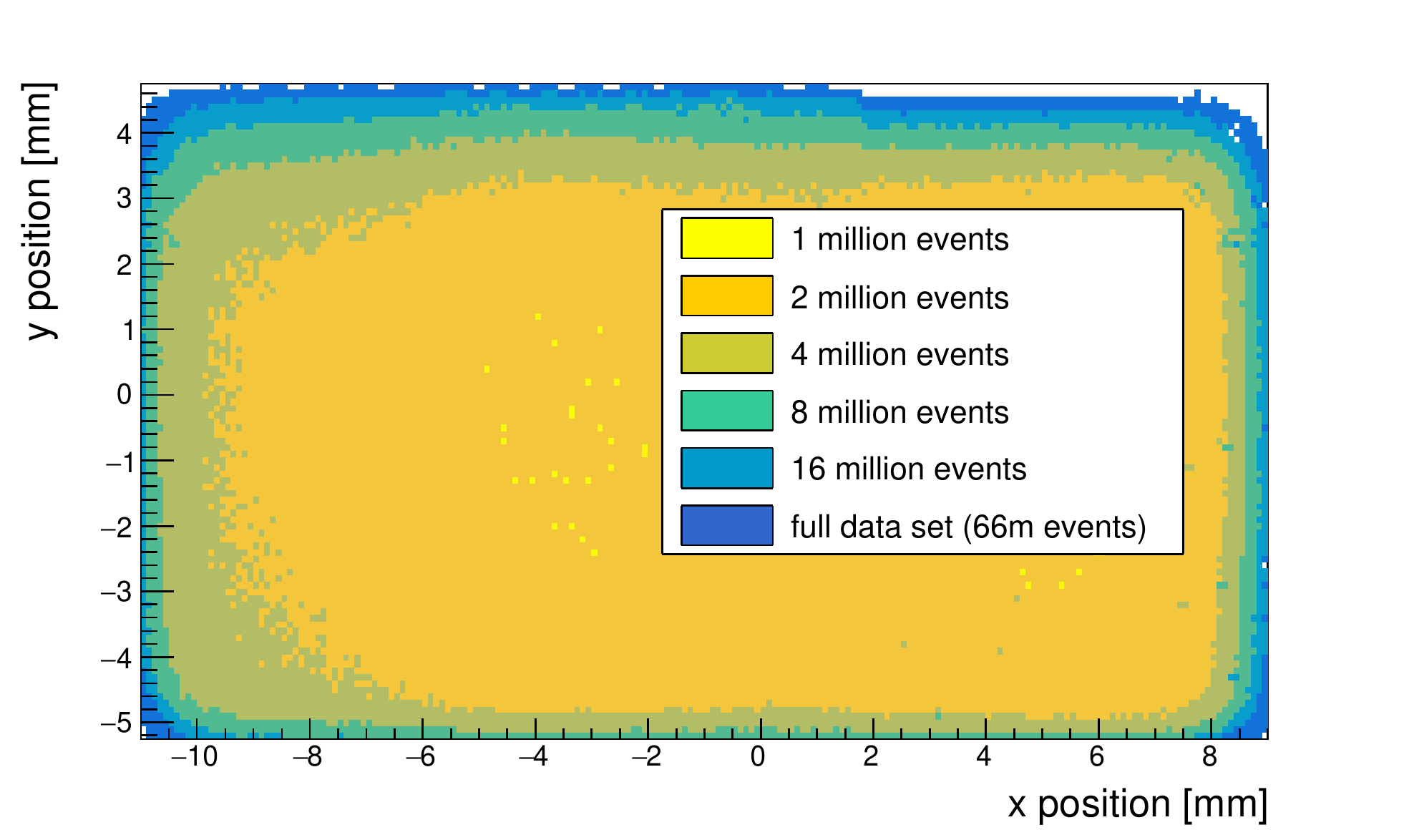}
  \caption{Illustration of the impact of the number of reconstructed events on the usable image area: for a minimum track requirement of 400 tracks, a minimum of 4 million events was found to result in a reasonably large reconstructed image area}
  \label{fig:nevents_1}
\end{subfigure}

 \begin{subfigure}{\textwidth}
  \centering
  \includegraphics[width=.9\linewidth]{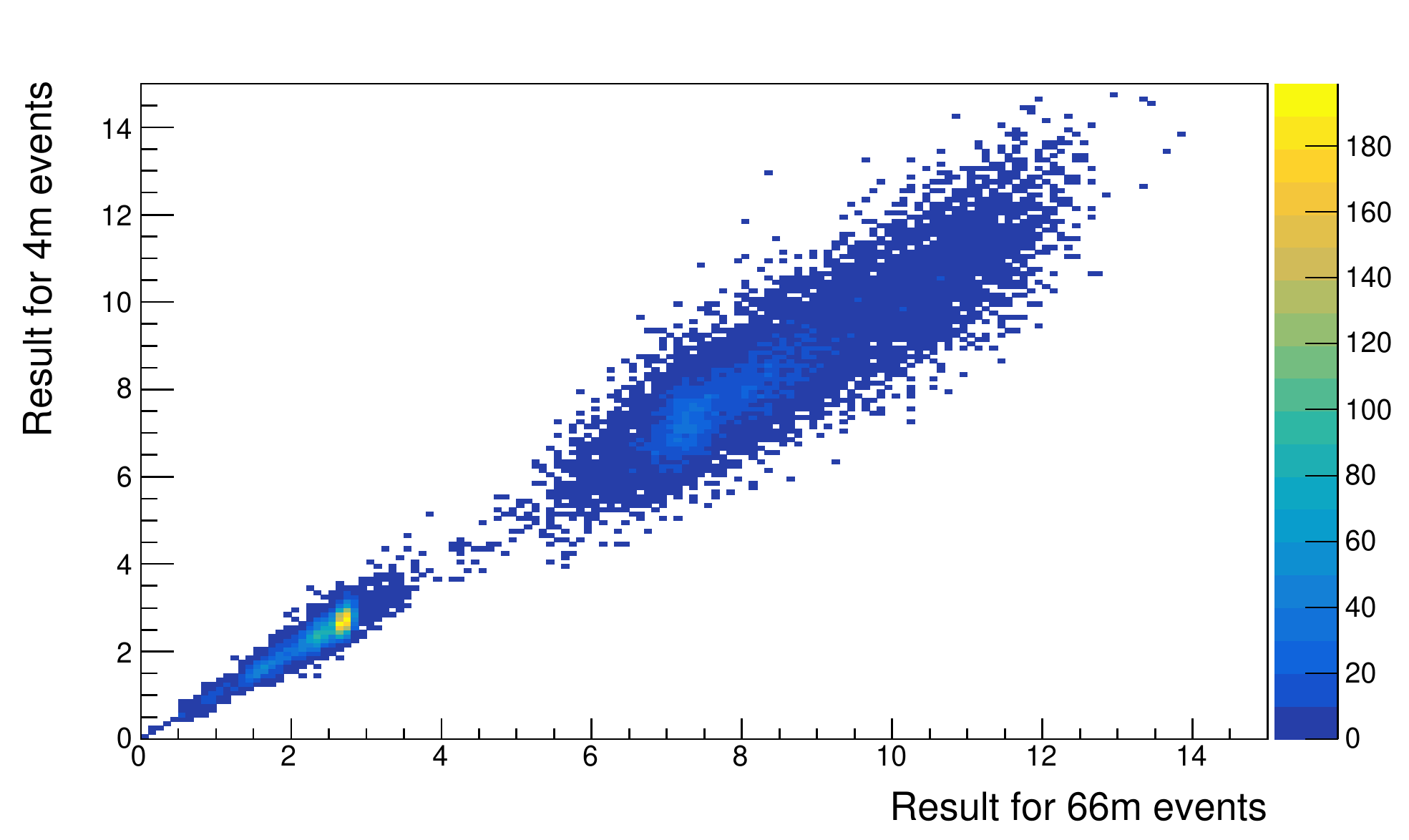}
  \caption{Correlation of the reconstructed material budget for an image reconstructed using 4 million and 66 million events. While low material budgets showed a good correlation for both data sets (correlation factors r = 0.93 for values $\unit[< 4.5]{\%\,X/X_0}$ and r = 0.89 for values $\unit[> 4.5]{\%\,X/X_0}$), large discrepancies were found for large material budgets, since small variations in the histogram can have large impacts on the fitted distribution width.}
  \label{fig:nevents_2}
\end{subfigure}
\caption{Impact of the size of the data set used for reconstruction on the reconstructed image area and pixel value.}
\end{figure}

This size was chosen as a compromise to maximise the reconstructed area per position (see figure~\ref{fig:nevents_1}) and precision of the reconstructed value (see figure~\ref{fig:nevents_2}), while minimising the beam time required per position.

\begin{figure}
  \centering
  \includegraphics[width=.9\linewidth]{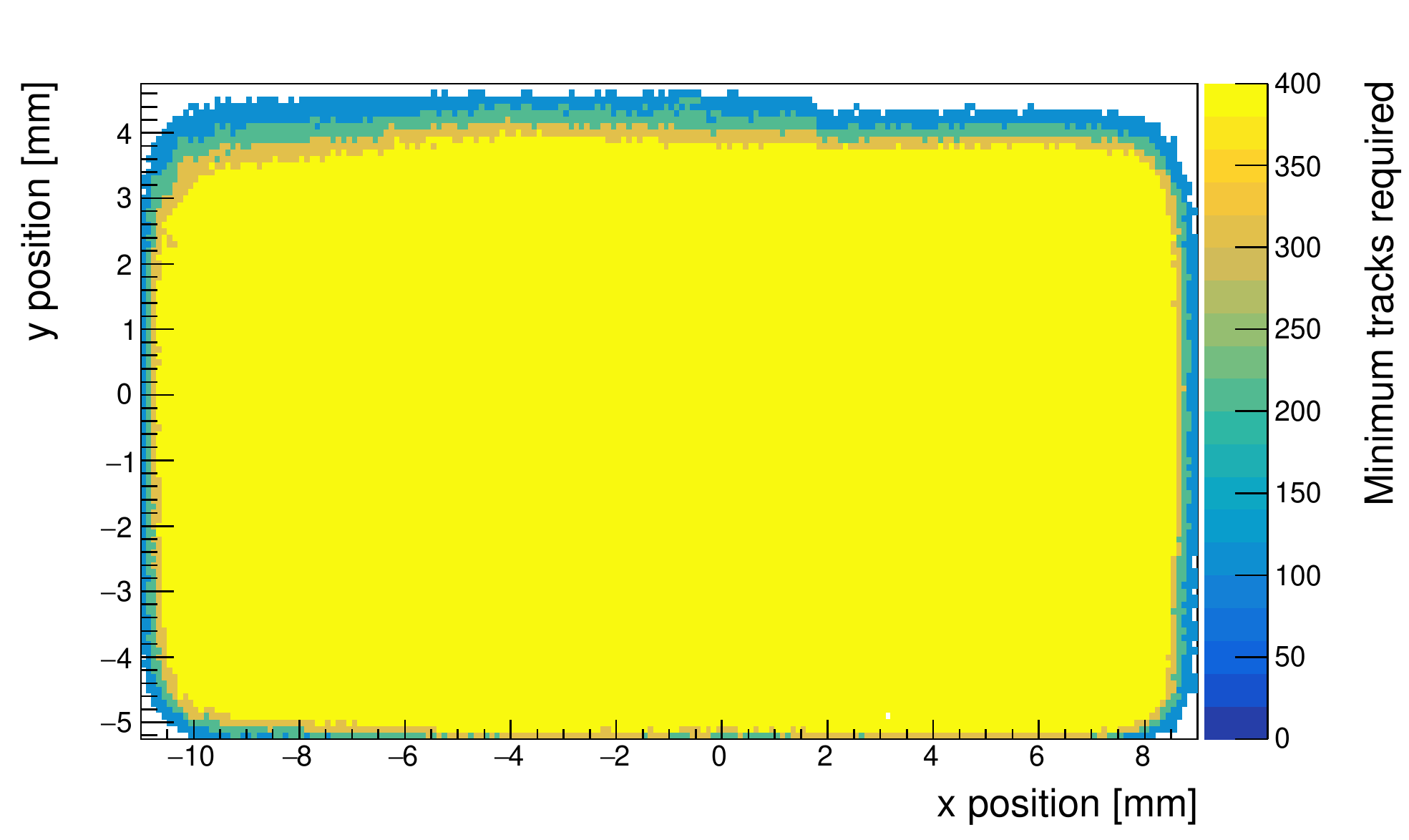}
  \caption{Image area that can be reconstructed from a 4 million data sample using different required minimum track cuts.}
  \label{fig:ntracks_1}
\end{figure}

Images were reconstructed requiring a minimum of 400 tracks per pixel, which was found to provide reliable results while leading to a sufficiently large reconstructed image area per position (see figure~\ref{fig:ntracks_1}.

\section{Results}
\label{sec:res}

\begin{figure}[htp]
  \centering
\begin{subfigure}{.48\textwidth}
  \includegraphics[width=\linewidth]{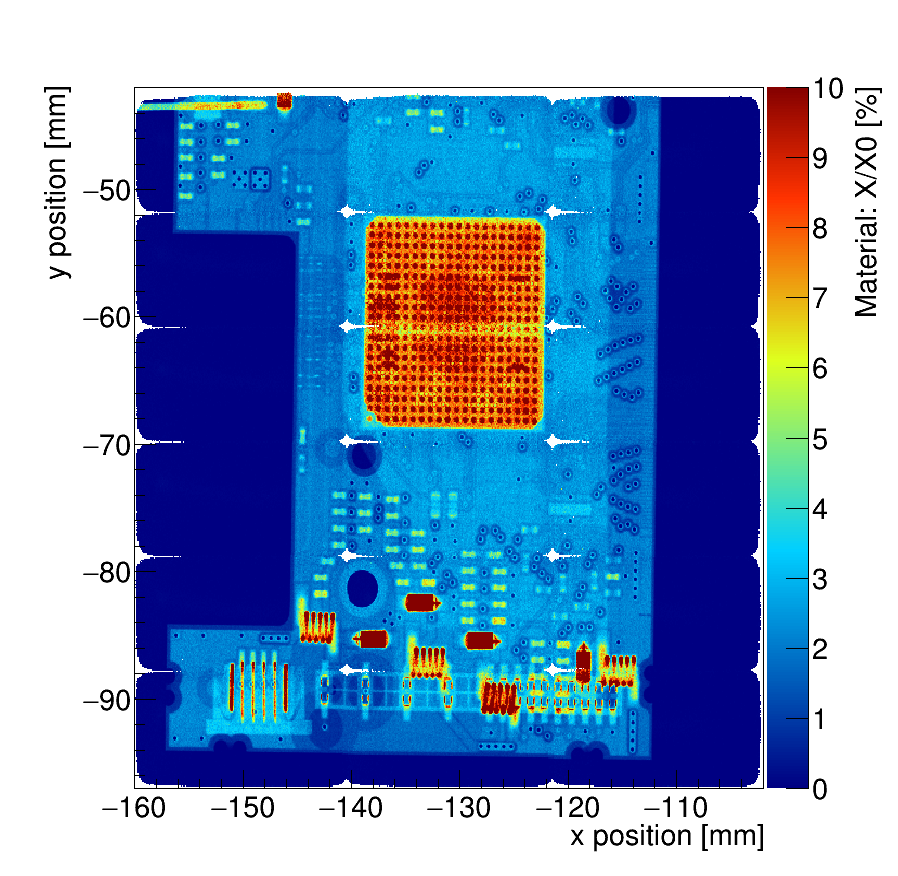}
  \caption{End of Structure card (top half) mapped in electron beam}
  \label{fig:EoS_map1a}
\end{subfigure}
\begin{subfigure}{.48\textwidth}
  \centering
  \includegraphics[width=\linewidth]{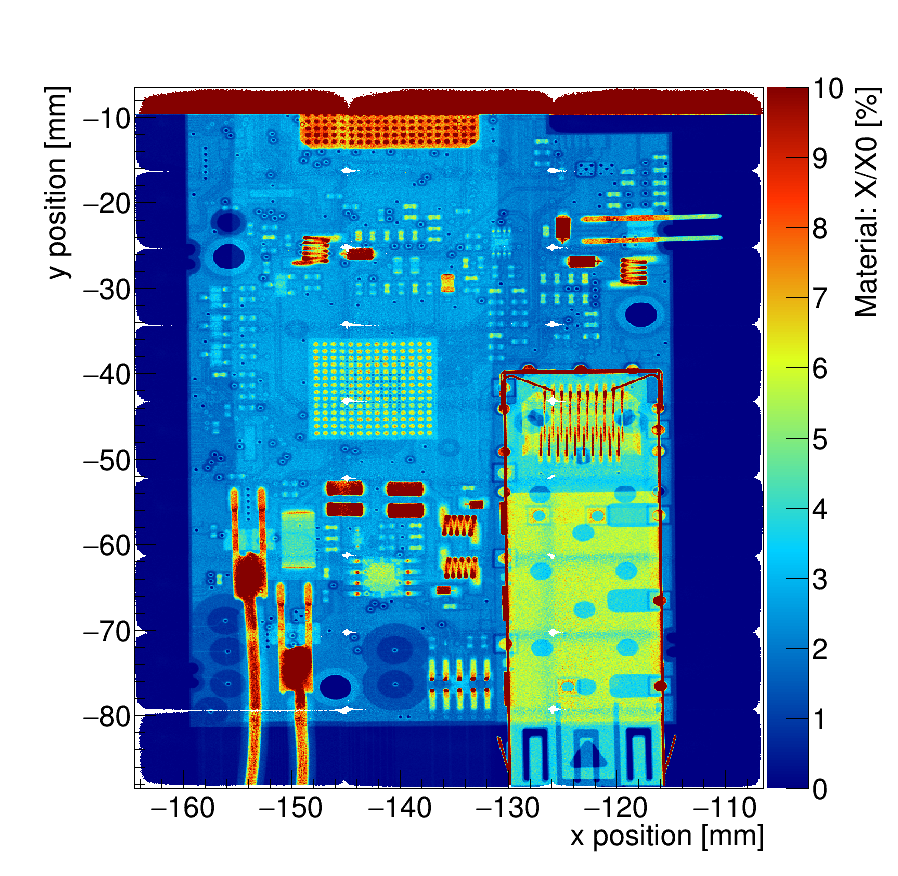}
  \caption{End of Structure card (bottom half) mapped in electron beam}
  \label{fig:EoS_map1b}
\end{subfigure}
\caption{Both sections of an EoS card mapped in electron beam in two sessions}
\label{fig:EoS_map1}
\end{figure}

Due to the size of the EoS card under investigation with respect to the support structure used to mount it in the electron beam, the EoS was mounted on the support structure in two different positions and mapped in two series of movements (see figures~\ref{fig:EoS_map1a} and~\ref{fig:EoS_map1b}).

\begin{figure}
    \centering
    \includegraphics[width=\linewidth]{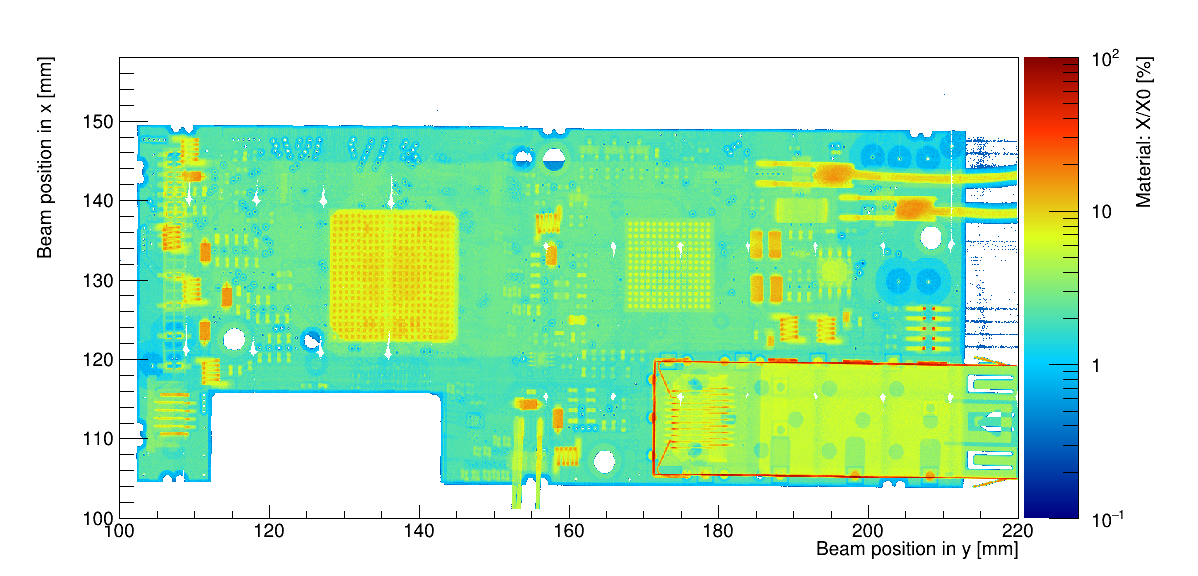}
    \caption{Full map of $X/X_0$ for an EoS card, calculated bremsstrahlung correction factor of $\epsilon = 0.3$, computed from two separate measurements. Combining both individual measurements into one overall EoS map showed a good agreement between both separate measurements.}
    \label{fig:EoS_combined_1}
\end{figure}

Both sections of the EoS card were afterwards joined using an algorithm that varied the planar rotation angle and displacement to find a maximum overlap between the common EoS regions on both maps (see figure~\ref{fig:EoS_combined_1}).
An improvement of the imaging algorithm was the implementation of a bremsstrahlung correction described above. By accounting for energy losses caused by bremsstrahlung, the resulting overestimation of the reconstructed material budget was compensated.

\begin{figure}
    \centering
    \includegraphics[width=0.9\linewidth]{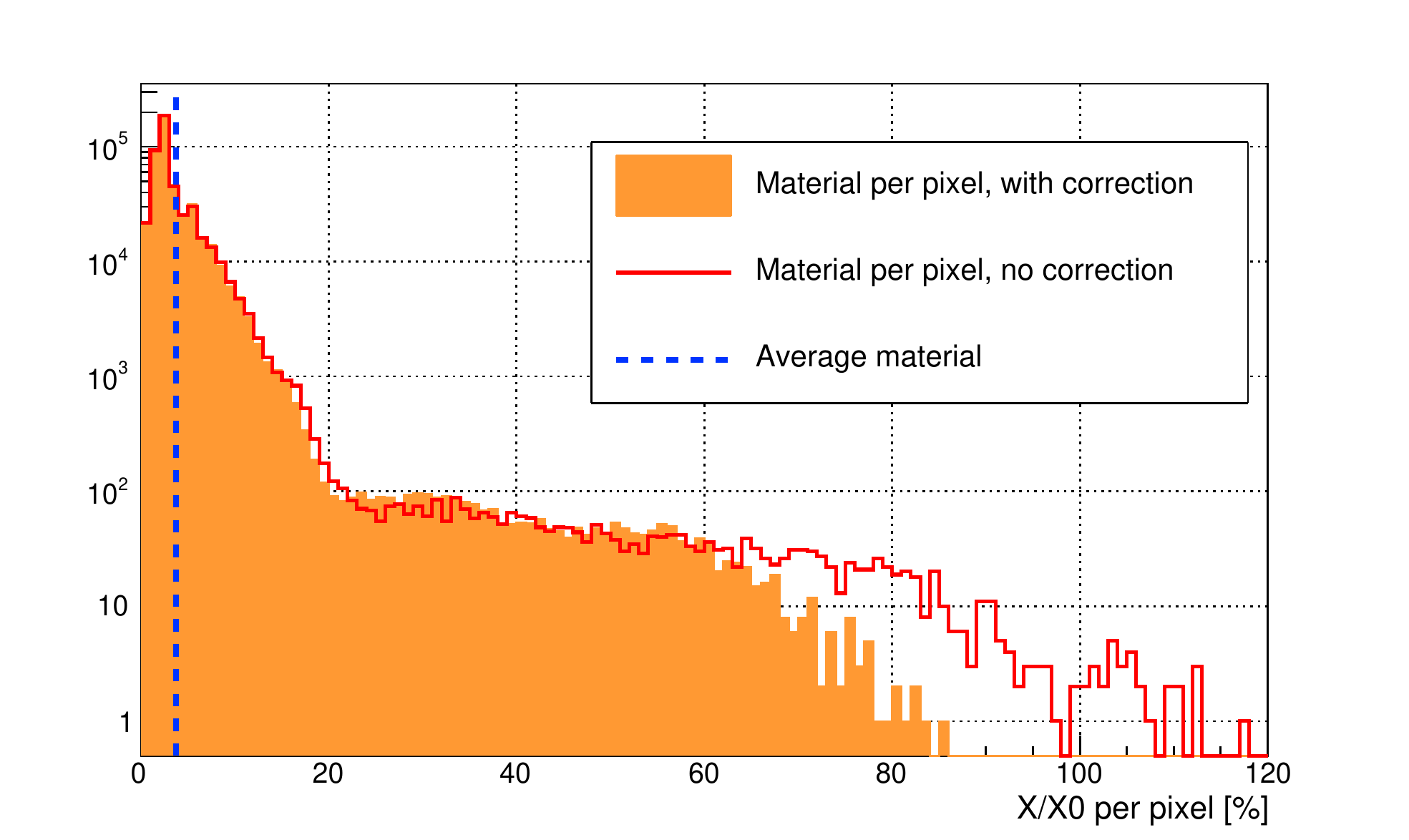}
    \caption{Distribution of reconstructed material per {$\unit[100\times100]{\upmu\text(m)^2}$} pixel for the full EoS card area (with and without correction). While the average amount of material is small, individual pixels show up to more than 20 times the estimated average (for values with bremsstrahlung correction).}
    \label{fig:perbin}
\end{figure}

\begin{figure}
    \centering
    \includegraphics[width=\linewidth]{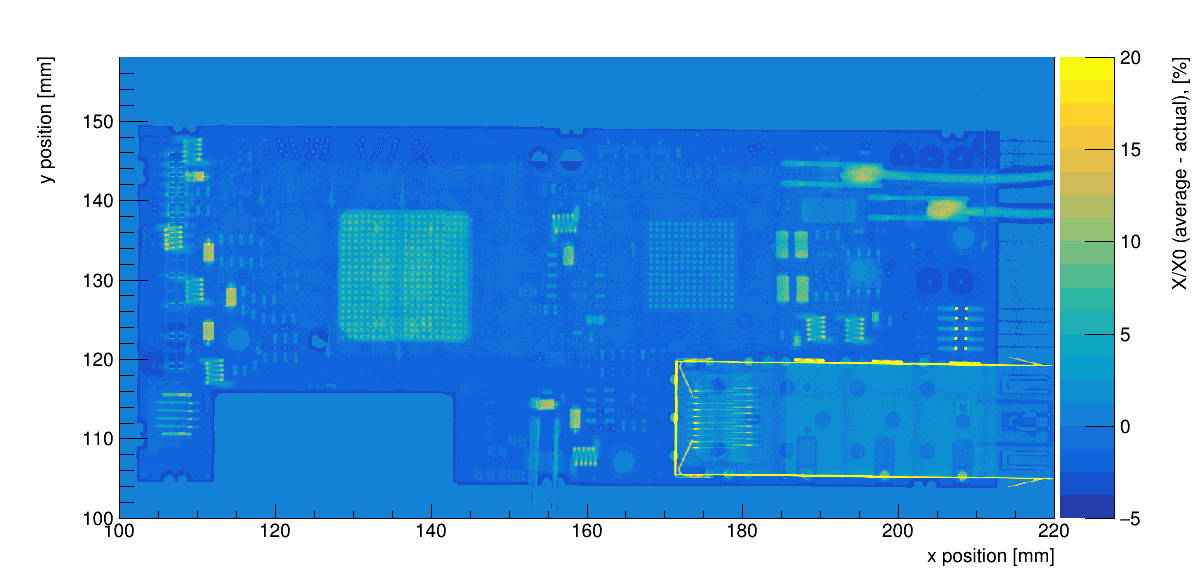}
    \caption{Map showing the local difference between an overall average and the actual material}
    \label{fig:EoS_diff}
\end{figure}

The obtained map was used for a comparison with the common approach of averaging all material over the full component area: calculating an average for the full EoS card results in \unit[3.7]{\% $X/X_{0}$}, assuming that no extra volumes are added for any of the components with more material, e.g. the copper cooling plate (see figure~\ref{fig:perbin}). 
This average can be used to calculate the discrepancy between average and actual amount of material per bin (see figure~\ref{fig:EoS_diff}).
The map shows several areas where material is underestimated, in addition to several components with significantly more material than estimated from the average.

\subsection{Comparison with known material}
\label{subsec:known}

In order to judge the quality of the reconstruction method, known features of the EoS card were used for comparison:
\begin{itemize}
    \item two pin headers with a diameter of \unit[0.64$\times$0.64]{mm$^2$}, made out of phosphor bronze, were mounted to extend over the edge of the EoS card (see figure~\ref{fig:pins1}) and could be used for a direct material comparison.
    \begin{figure}[htp]
    \centering
    \begin{subfigure}{.8\textwidth}
    \includegraphics[width=\linewidth]{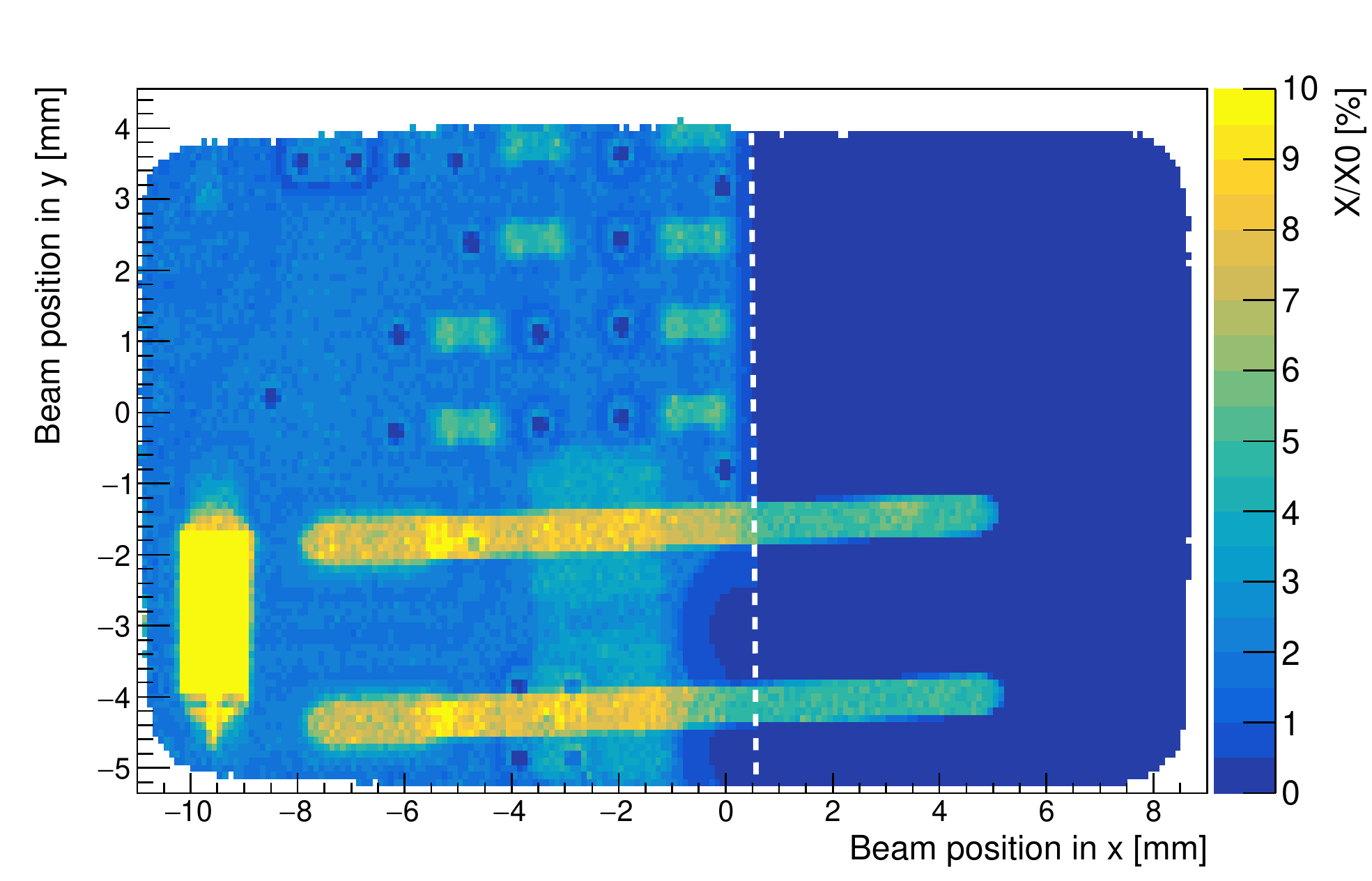}
    \caption{Region of EoS card with two pins extending over the edge of the EoS (edge of the PCB marked in white)}
    \label{fig:pins1}
    \end{subfigure}
    \begin{subfigure}{.8\textwidth}
    \centering
    \includegraphics[width=\linewidth]{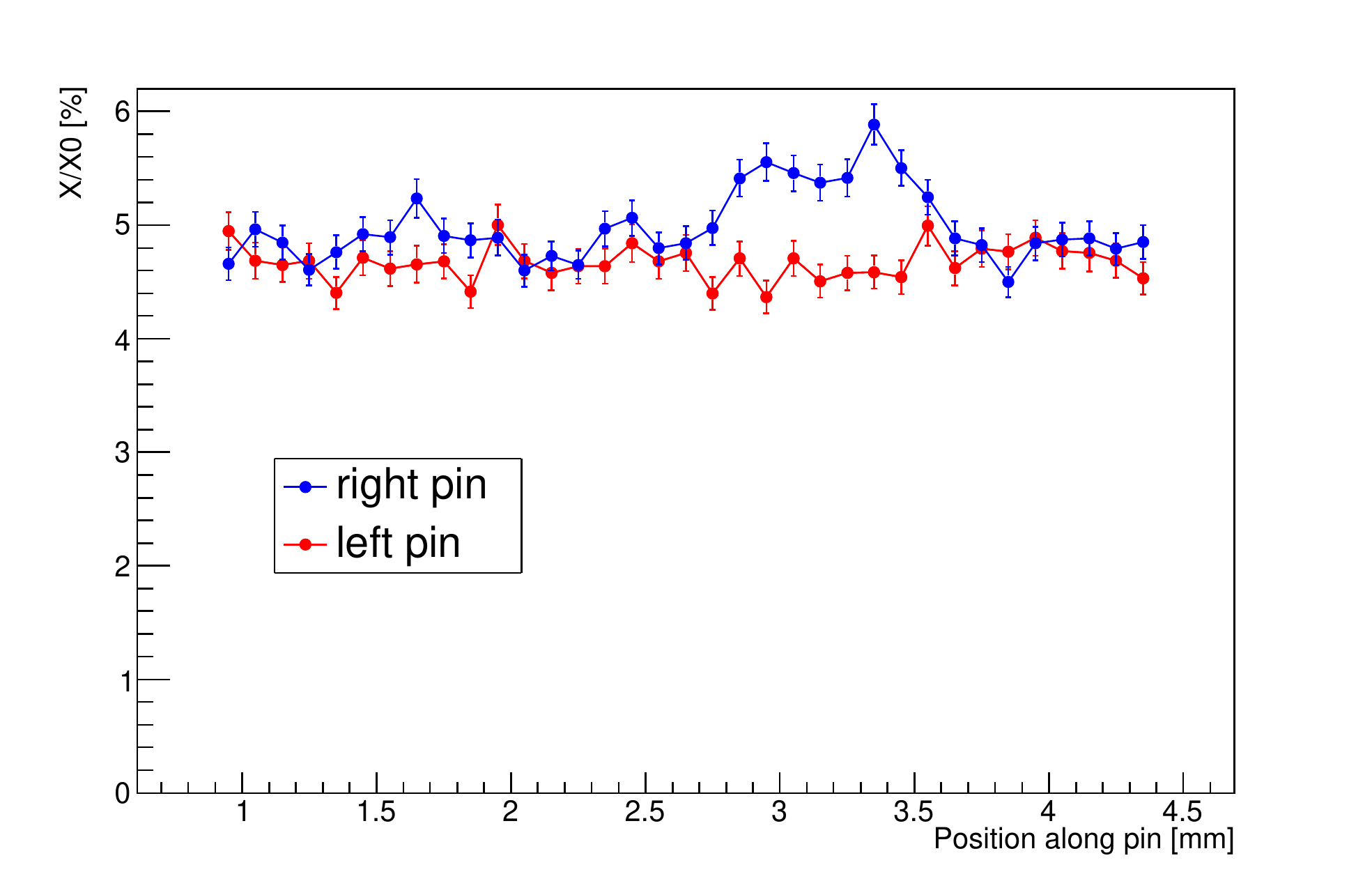}
    \caption{Average material per vertical scan line per pin, calculated with bremsstrahlung correction}
    \label{fig:pins2}
    \end{subfigure}
    \caption{Analysis of reconstructed material for two pin headers extending over the edge of an EoS card, i.e. with no background material except for air. The pin headers were treated as homogeneous (except for some solder residue on the right pin), which agreed well with the estimated values.}
    \label{fig:pins}
    \end{figure}
    While the exact composition of the bronze pins is unknown, it can be estimated as a standard bronze alloy consisting of \unit[60]{\%} copper ($X_0 = \unit[14.36]{\text{mm}})$ and \unit[40]{\%} tin ($X_0 = \unit[12.06]{\text{mm}})$, leading to a combined radiation length of \unit[13.44]{mm}. With a thickness of \unit[0.64]{mm}, each pin can be expected to correspond to a material contribution of \unit[4.8]{\% $X/X_0$}, which agrees well with the estimated values within uncertainties. Additionally, both pins are consistent with each other (except for an area with solder residue on one pin) and no gradient is observed over the length of the pin, which indicates a correctly determined beam energy calibration factor.
    
    \item Two wires with a cross-section of \unit[1]{mm$^2$} can be seen leading away from the EoS card (see figure~\ref{fig:wire_area}).
    \begin{figure}
    \centering
    \includegraphics[width=0.7\linewidth]{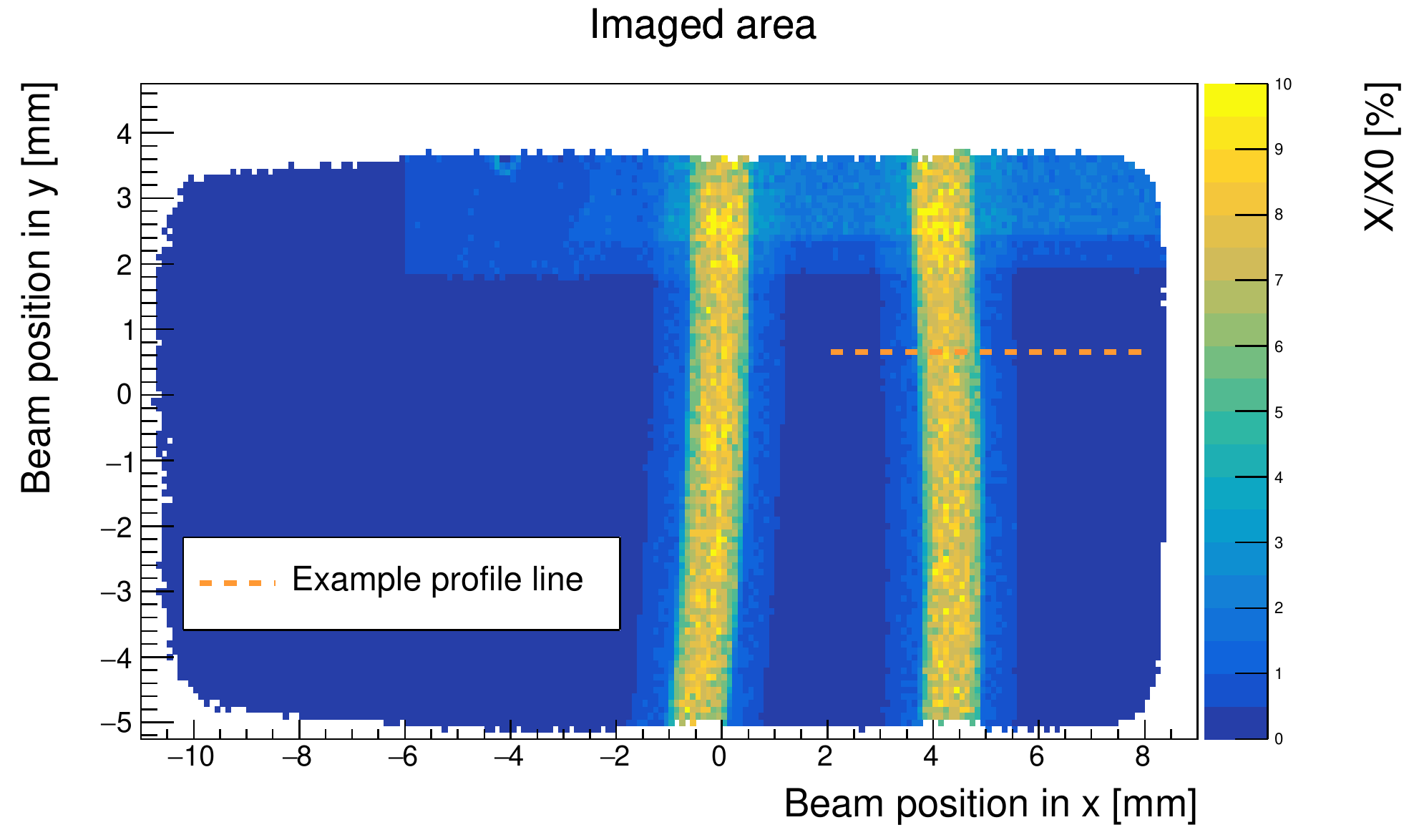}
    \caption{Mapped image area showing two wires with insulation leading away from the EoS card.}
    \label{fig:wire_area}
    \end{figure}
    Due to the round cross-section of the cable, a fit function was used to calculate the radiation length of both the metal core of the cable and the insulation material (see figure~\ref{fig:wire_fit}).
    \begin{figure}
    \centering
    \includegraphics[width=0.7\linewidth]{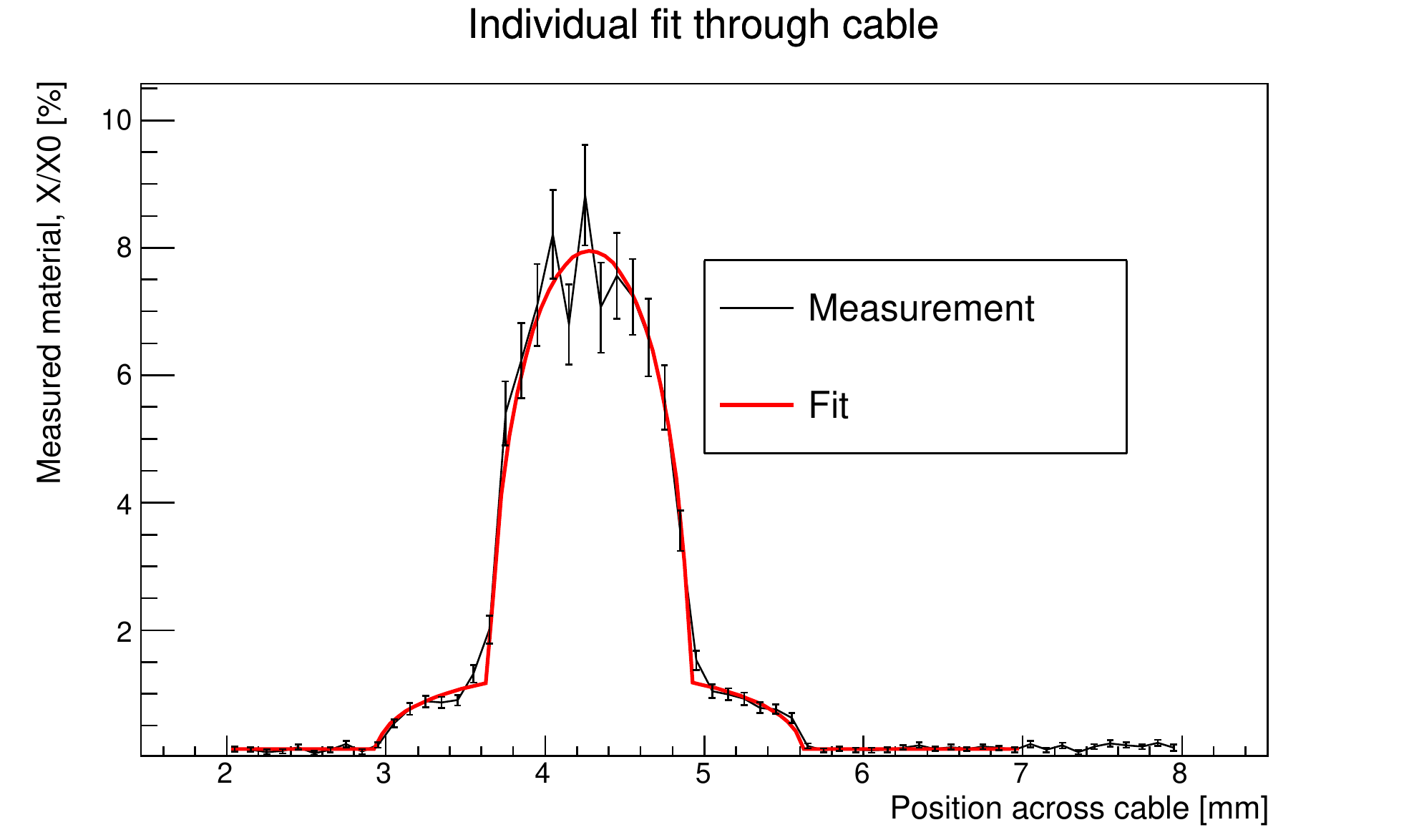}
    \caption{Reconstructed material along one line across an individual cable and applied fit function: the low material regions correspond to image areas with insulation only, the centre shows the metal core of the cable.}
    \label{fig:wire_fit}
    \end{figure}
    The fit results were used to calculate the radiation length X$_{0}$ of both the metal and the insulation comprising both cables (see figures~\ref{fig:wires1} and~\ref{fig:wires2}). The determined radiation lengths show good agreement along each cable and between the two cables, which were of the same type. While the exact material contributions of both cables are unknown, the determined values are in good agreement with a metal on the inside and a plastic on the outside of the cable.
    \begin{figure}[htp]
    \centering
    \begin{subfigure}{.7\textwidth}
    \includegraphics[width=\linewidth]{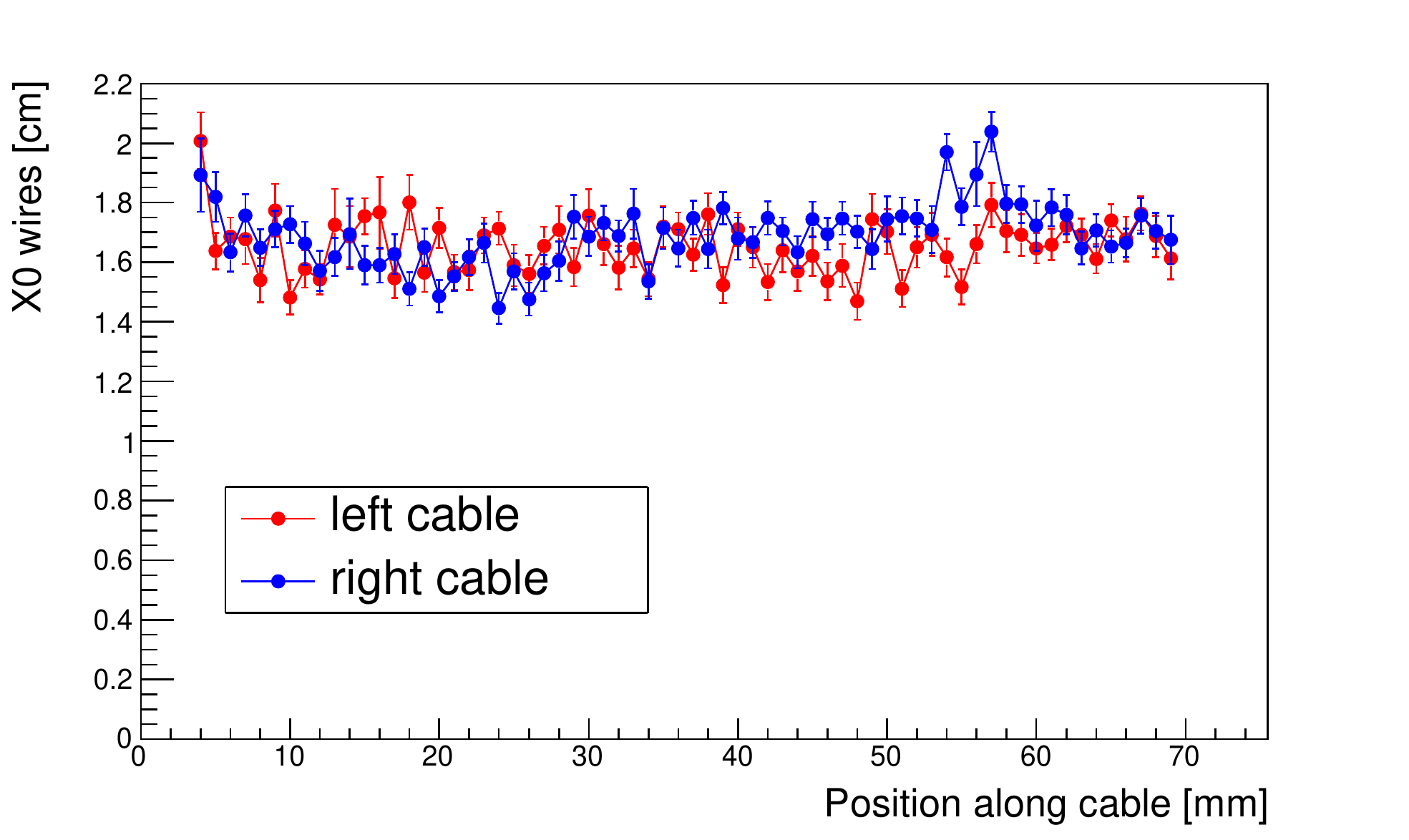}
    \caption{Radiation length of metal cable core, determined from fit, for left and right cable (determined with applied bremsstrahlung correction)}
    \label{fig:wires1}
    \end{subfigure}
    \begin{subfigure}{.7\textwidth}
    \includegraphics[width=\linewidth]{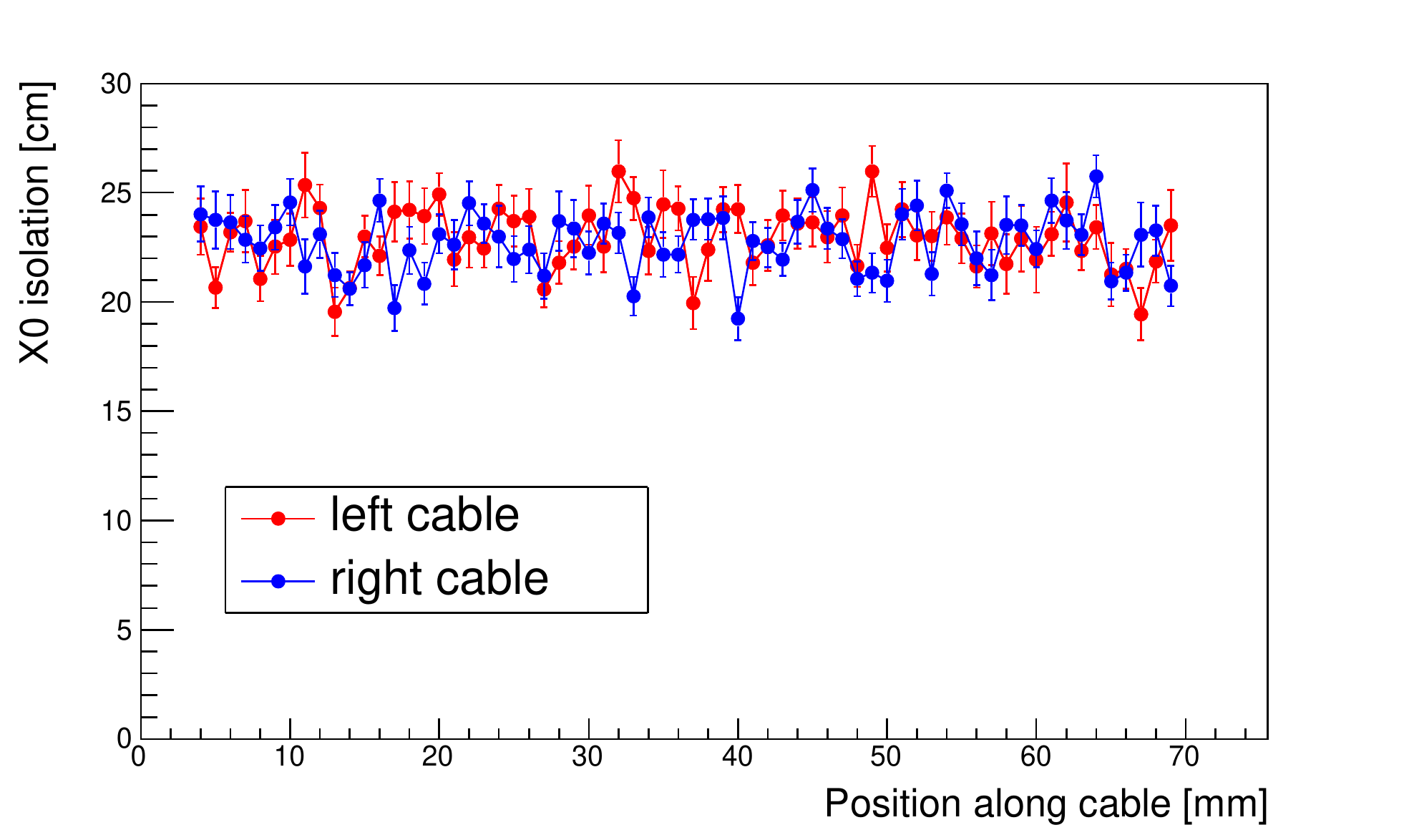}
    \caption{Radiation length of cable insulation material, determined from fit, for left and right cable (determined with applied bremsstrahlung correction)}
    \label{fig:wires2}
    \end{subfigure}
    \caption{Radiation lengths of cable materials determined per image line. The results are consistent within uncertainties within uncertainties for both cables along the cable length.}
    \label{fig:wires}
    \end{figure}

    \item drill holes with vias can be seen in several areas of the EoS card: they traverse all layers of the EoS, with a \unit[20]{$\upmu$m} copper layer on the inside of the drill hole for electrical contact (see figure~\ref{fig:EoS_Stack}). For a comparison with the expected material, a drill hole was chosen which was not covered by the protective layer on the EoS card (see figure~\ref{fig:EoS_foto}). A picture of the hole used for analysis is shown in figure~\ref{fig:smallhole1}.
    \begin{figure}[htp]
    \centering
    \includegraphics[width=0.8\linewidth]{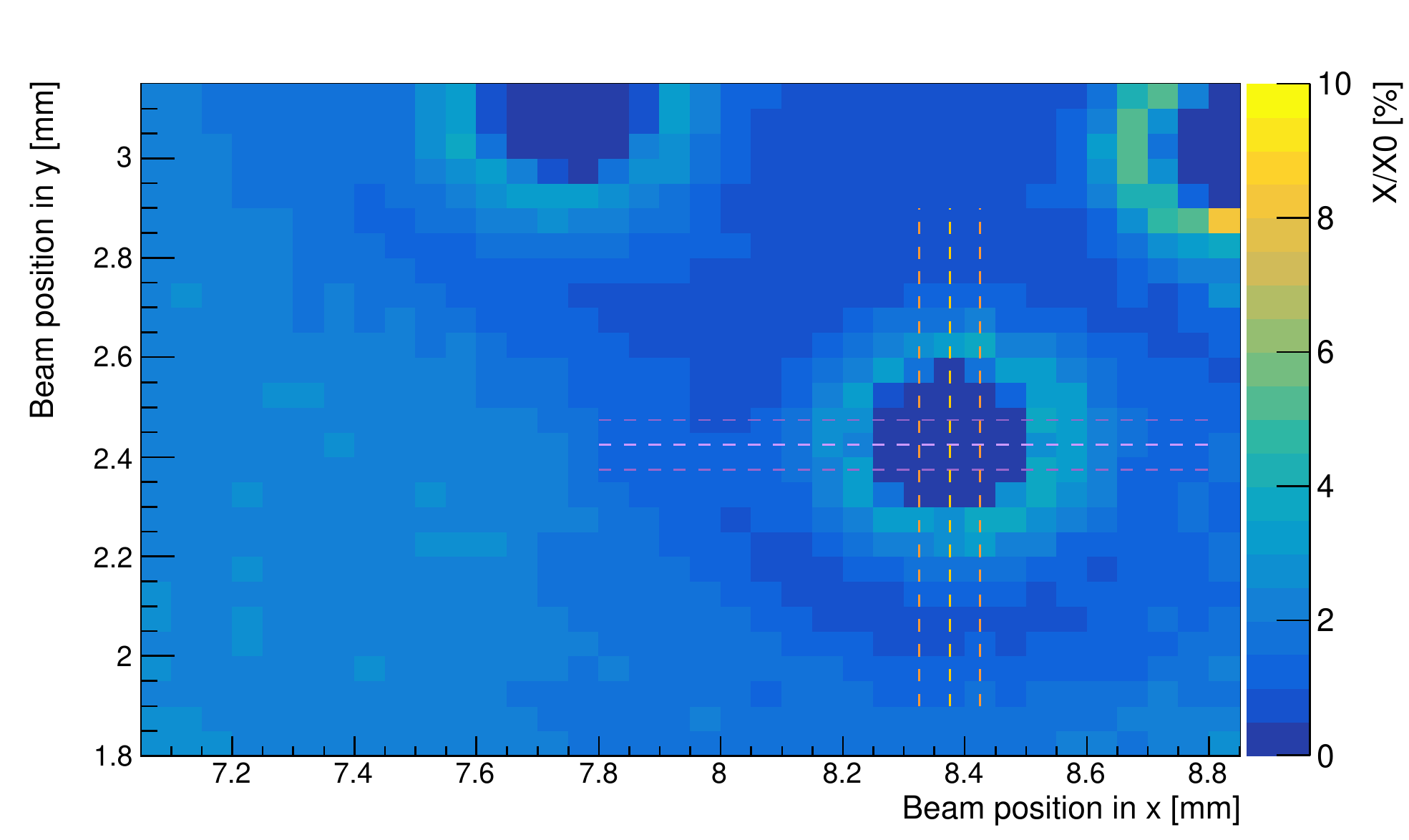}
    \caption{Detail area of an EoS imaging scan showing a drill hole with a vias in an unobstructed area of the EoS card. Dashed lines indicate positions used to analyse the material distribution of the drill hole (see figure~\ref{fig:smallhole2}).}
    \label{fig:smallhole1}
    \end{figure}
    As shown in figure~\ref{fig:EoS_Stack}, an EoS FR4 stackup without any copper amounts to approximately \unit[0.75]{\%} $X/X_{0}$, which is expected for the material surrounding the drill hole. Since there is no material within the hole, the reconstructed material is expected to be zero (within uncertainties) in that area. Figure~\ref{fig:smallhole2} shows the obtained material distribution.
    \begin{figure}[htp]
    \centering
    \includegraphics[width=0.8\linewidth]{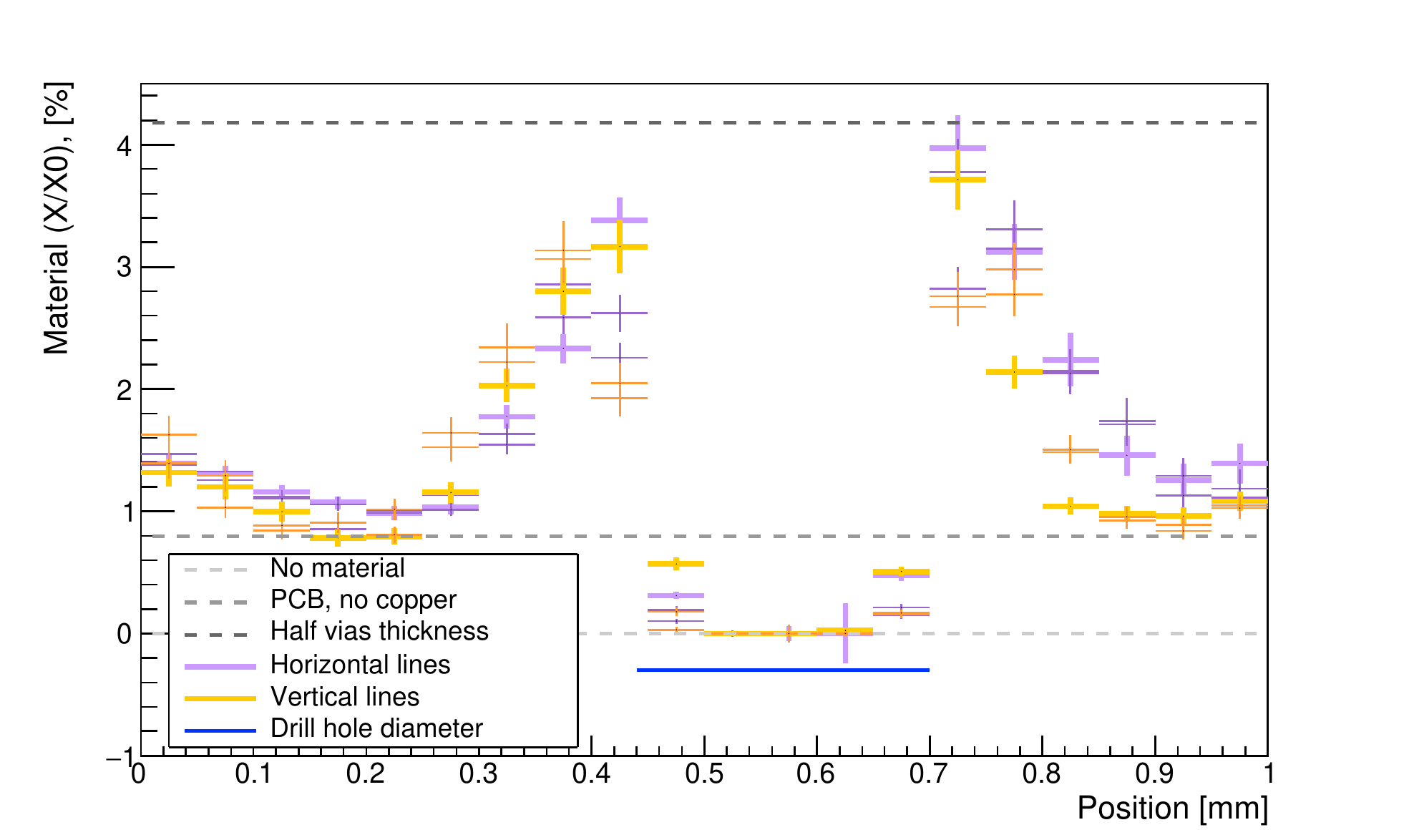}
    \caption{Material profile along horizontal and vertical profile lines positioned around a drill hole with vias.}
    \label{fig:smallhole2}
    \end{figure}
    It was found that the calculated values agree well with the expectations: in the absence of copper traces in the PCB, the reconstructed material reaches a value of \unit[0.75]{\%} $X/X_{0}$, as well as reconstructed material consistent with zero (within uncertainties) at the centre of the drill hole. The ring surrounding the drill hole was found to be significantly lower than expected from the copper wall inside the vias, which corresponds to \unit[1.2]{mm} of copper (\unit[8.4]{\%} $X/X_{0}$). It has a thickness of about \unit[20]{$\upmu$m} or less than half a pixel length, which leads to a combined value for PCB and vias areas.
    \begin{figure}[htp]
    \centering
    \includegraphics[width=0.8\linewidth]{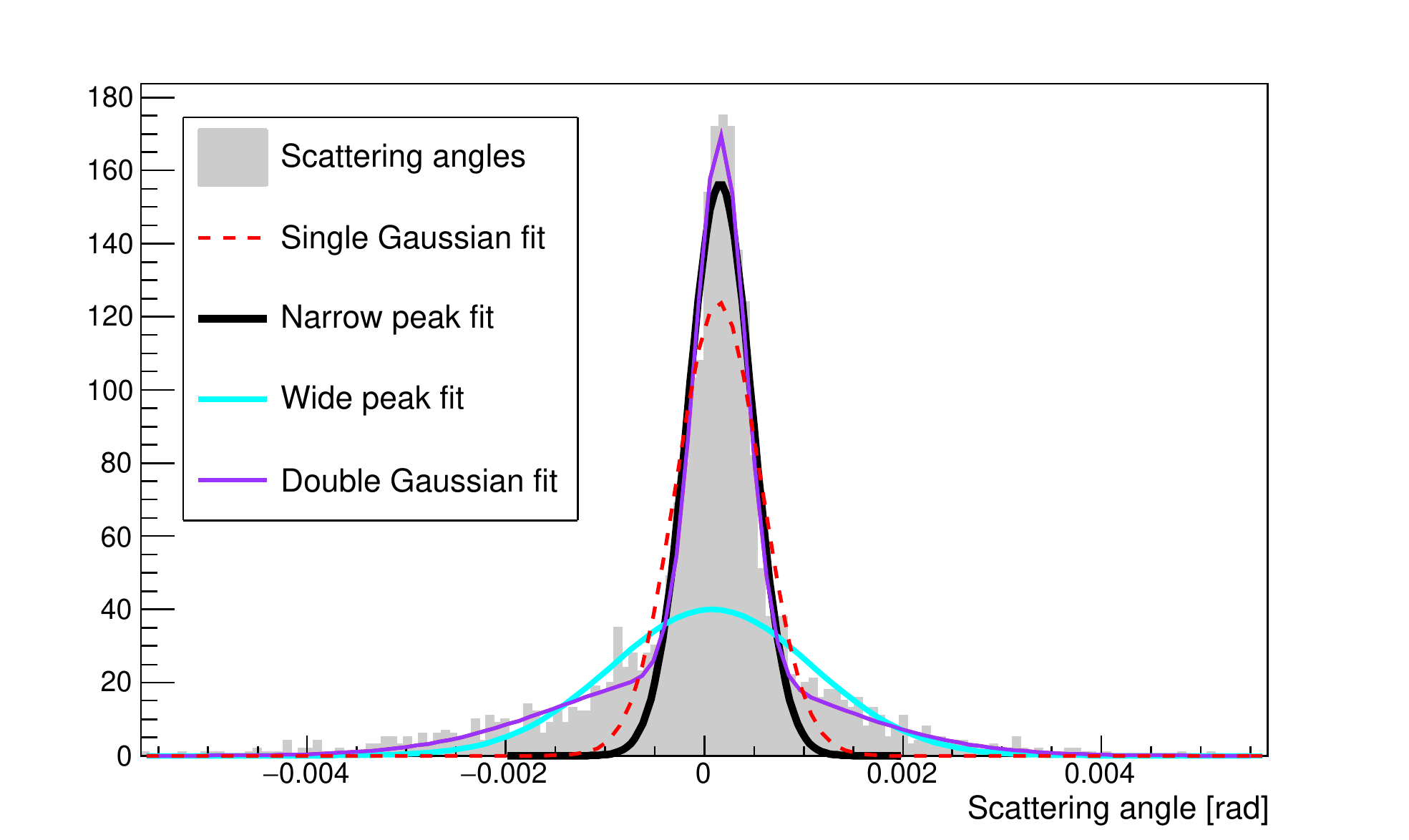}
    \caption{Scattering angle distribution from a pixel at the edge of a vias showing the contributions from areas with little material and significantly more scattering material. While the standard approach of a single Gaussian fit (dashed red line) does provide a reasonable approximation of the narrow peak (corresponding to air), it underestimates the impact of the wider peak. In a second approach, the two peaks were first fitted separately (black line fitting the narrow peak (air), cyan line corresponding to the wide peak (copper-plated wall of the vias)) and the obtained fit parameters used as input parameters to a double Gaussian fit (violet line), which shows a better approximation of the scattering angle distribution.}
    \label{fig:failed_vias}
    \end{figure}
    In figure~\ref{fig:failed_vias}, this effect is demonstrated: a single Gaussian fit results in an overall material estimate of \unit[0.5]{\% $X/X_{0}$}. Fitting both peaks separately yields two distinct Gaussians: a narrow peak (\unit[0.01]{\% $X/X_{0}$}, consistent with air) and a wide peak (\unit[5.7]{\% $X/X_{0}$}, which is lower than expected for \unit[1.2]{mm} of copper, but can be expected due to the thin wall thickness and large scattering angles, which could scatter the particles out of the copper wall). Weighing them based on the areas under the respective curves leads to a combined value of \unit[2.5]{\% $X/X_{0}$}, which is distinctly larger than the estimate based on a single Gaussian fit and, judging from the distributions, a more reasonable approximation.
\end{itemize}

Studies of dedicated areas on the EoS card found a good agreement between reconstructed and known material, where available.

\subsection{Pixel size}

The resolution that can be achieved from electrons is determined by several factors:
\begin{itemize}
    \item the intrinsic resolution of the beam telescope~\cite{Rubinskiy2012923}
    \item offsets in the reconstructed tracks which can occur for thick layers of materials with high density
    \item the minimum pixel size that can be used for the image reconstruction process.
\end{itemize}
The image pixel size is usually determined by the size of the data sample: since a minimum number of tracks is required to perform a reliable fit, the overall image needs to be divided into pixels containing the required minimum of tracks. Scattering angles from different materials within the same pixel lead to a smeared Gaussian distribution. Applying a single Gaussian fit function to the resulting angle distribution can naively be assumed to result in a reconstructed scattering angle width corresponding to an average of the different material contributions.

\begin{figure}[htp]
    \centering
    \includegraphics[width=\linewidth]{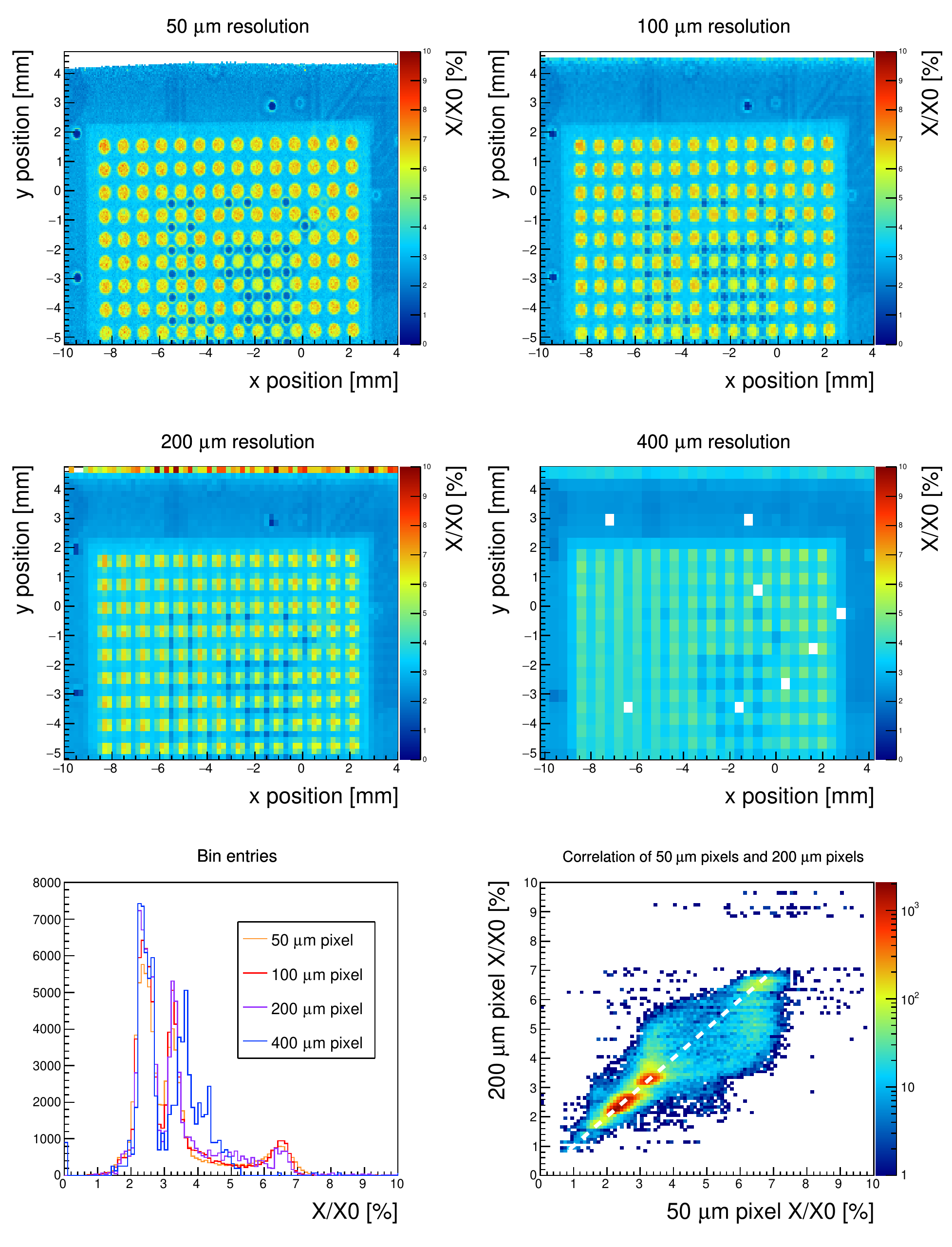}
    \caption{Image reconstructed for a readout chip with a cooling plate for different pixel sizes. A correlation plot for the same area reconstructed using different pixel sizes shows that, for this area, a larger pixel size can cause a systematic shift towards lower values of material.}
    \label{fig:pixsize1}
\end{figure}

This assumption was tested in these measurements by collecting a large data set for an area under investigation that was known to have an uneven material distribution: a readout chip with a heat sink. A data set of 56 million events was obtained to provide a sufficient number of tracks per pixel to ensure a reliable fit and the same data sample was reconstructed using pixel sizes of \unit[50, 100, 200 and 400]{$\upmu$m} (see figure~\ref{fig:pixsize1})

\begin{figure}[htp]
    \centering
    \includegraphics[width=0.8\linewidth]{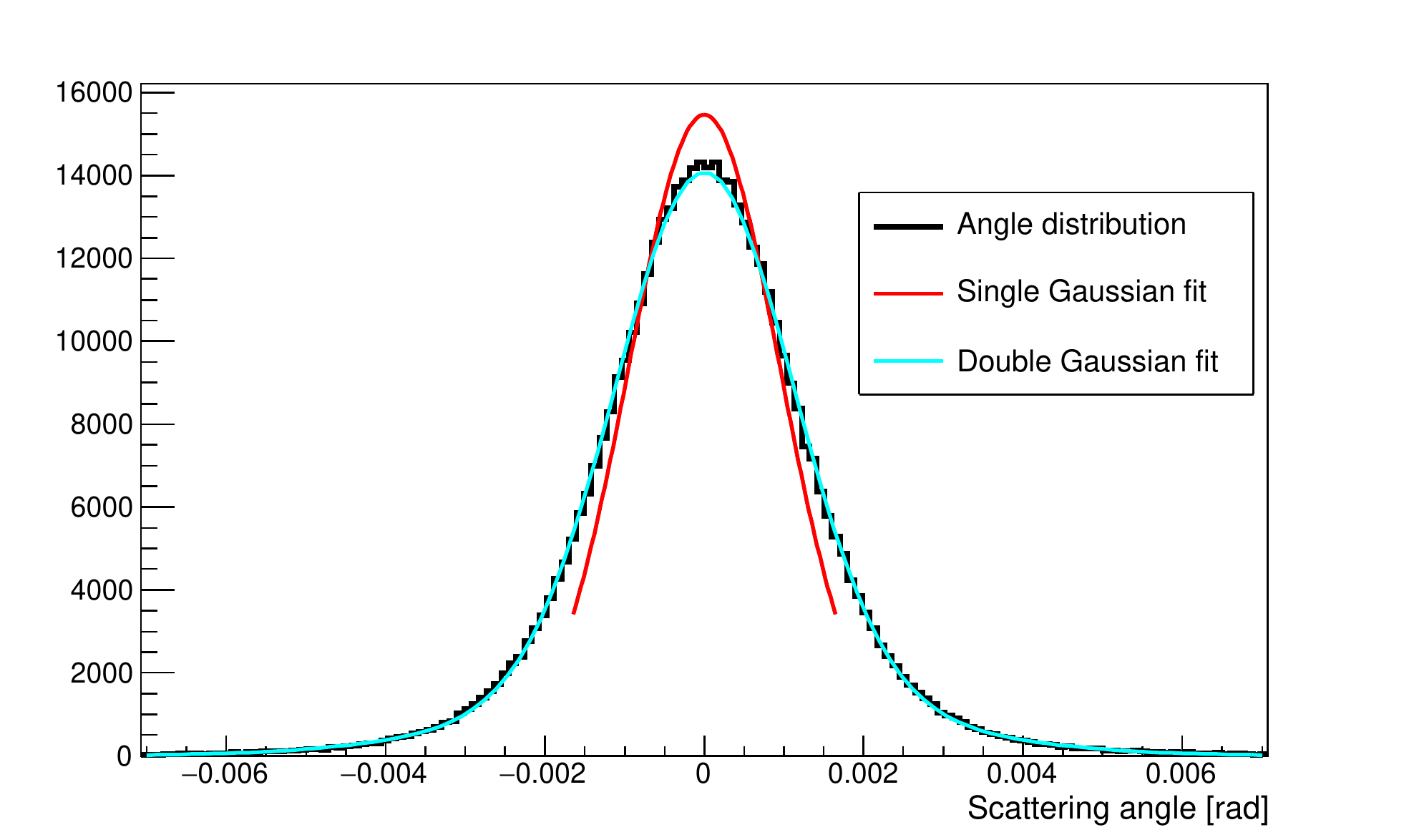}
    \caption{Example of a scattering angle distribution that can not be described by a single Gaussian fit: due to an inhomogeneous material distribution within a pixel, the scattering angle distribution consists of an overlay of Gaussian distributions and can not be approximated by a Gaussian fit function. Using a combination of two independent Gaussian distributions (cyan line) improves the description of the observed scattering angle distribution.}
    \label{fig:failedfit}
\end{figure}

Comparing the reconstructed images showed that up to a pixel size of \unit[200]{$\upmu$m}, the maximum and minimum reconstructed material budget values were comparable for all images. For a pixel size of \unit[400]{$\upmu$m} the reconstructed image was found to lead to an overall reduction of the average material budget.
Additionally, a pixel size of \unit[400]{$\upmu$m} was found to result in pixels where the obtained track angle distribution could not be described by the applied fit function anymore and the fit failed (see figure~\ref{fig:failedfit}). 

\begin{figure}[htp]
    \centering
    \includegraphics[width=\linewidth]{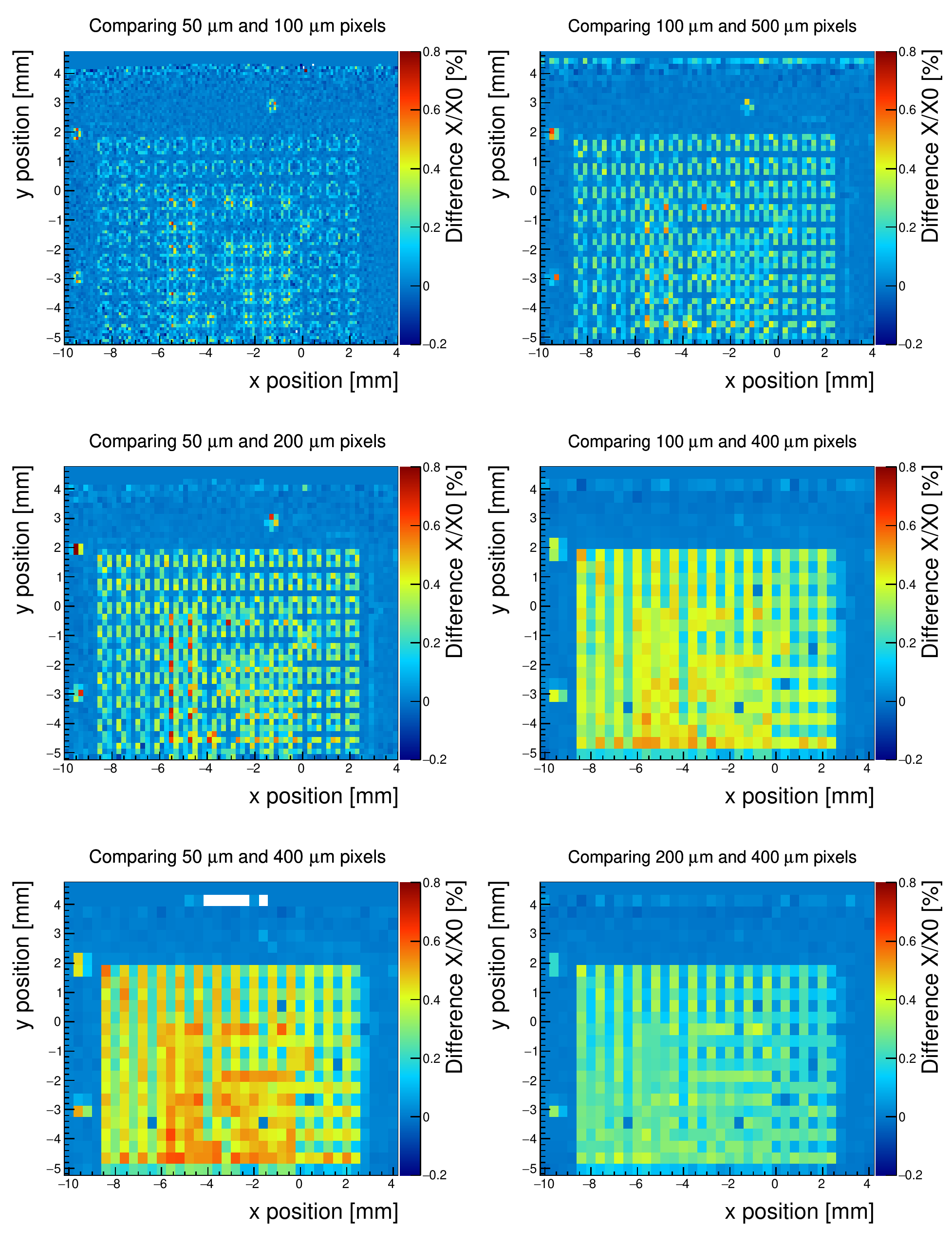}
    \caption{Deviation maps of images constructed by averaging smaller pixels with the corresponding image reconstructed with a larger pixel size}
    \label{fig:pixsize2}
\end{figure}

For a quantifiable comparison of the impact of larger pixel sizes, images with retroactively increased pixel sizes were constructed for comparison with their reconstructed equivalent:
matrices of smaller pixels were averages into larger pixels and the resulting image was compared with an image reconstructed with that same pixel size (see figure~\ref{fig:pixsize2}).

The resulting maps showed that the pixel size used for reconstruction has little impact on the reconstructed material budget for homogeneous material, but does affect the reconstructed value in very inhomogeneous regions. For the region under investigation, the size of individual features was about \unit[200]{$\upmu$m}, and the obtained maps showed only localised discrepancies for pixel sizes up to that size, with only small and localised discrepancies between maps reconstructed using pixel sizes of \unit[50]{$\upmu$m} and \unit[100]{$\upmu$m}. However, larger pixel sizes show systematically lower material budgets than smaller pixel sizes, which affect the overall reconstructed average and lead to a systematically lower reconstructed value.

Studies with random number generated track angles show that in cases of pixels with different fractions of material within the pixel, fitting the resulting superposition of Gaussian distribution leads to a calculated distribution width that is dominated by the biggest fraction of material contained in the pixel rather than the expected combination of both values.

Since the EoS region shown here has a mostly low material budget with pillars of more material occupying a smaller fraction of space, pixels can be assumed to frequently have smaller fractions of large material areas and therefore to disproportionately reflect the lower material budget areas, which leads to a general underestimation. Studying this area led to the conclusions that in general, pixel sizes for image reconstruction should be chosen to be smaller than features on the imaged object in order to avoid the surrounding area to dominate the obtained material budget.

Additionally, work was started on a method to account for uneven material distributions within one pixel by fitting track angle distributions with a mixture of two Gaussian densities and using their widths to calculate a weighted mean (see figure~\ref{fig:failedfit}) rather than a single Gaussian fit (as defined in the algorithm used here, see section~\ref{sec:radiationlength_measurements}), which is planned to be included in future versions of the reconstruction algorithm.

\section{Impact on Tracking}

After reconstructing the material distribution for the EoS card under investigation, the obtained map was used to study the impact of the material distribution on the tracking performance. For this study, a subset of 1 million events in one area of the EoS card with a range of material thicknesses was used (image area with pin headers, see figure~\ref{fig:pins1}). Tracks were reconstructed for three scenarios:
\begin{itemize}
    \item air, where no material was estimated for the EoS card layer
    \item constant, where a constant, flat material density of \unit[3.7]{\%} was assumed for the whole EoS area, corresponding to the calculated average over the full EoS area
    \item map, where the estimated material distribution was used for the reconstruction
\end{itemize}

For each scenario, a straight line fit of tracks was applied, where both multiple scattering and energy loss were taken into account. The straight line fit was implemented using a Kalman filter starting from a hit on the first telescope plane. In the absence of a magnetic field, the momentum affects the track fit only via the multiple scattering variance from the Highland model. The initial momentum at the first plane was set to the beam energy and subsequently decreased by the mean energy loss caused by bremsstrahlung depending on the traversed material.

\begin{figure}[htp]
    \centering
    \includegraphics[width=0.8\linewidth]{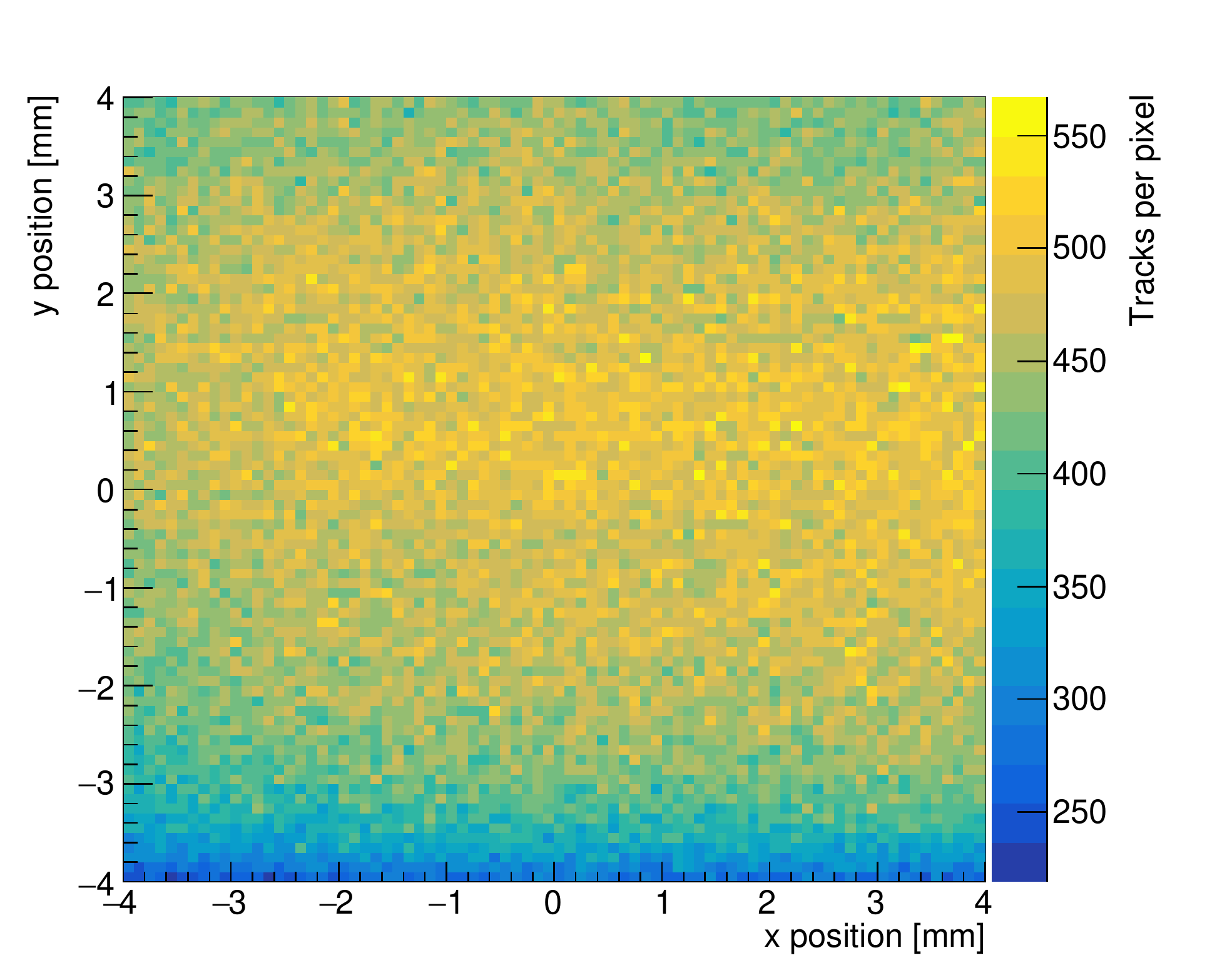}
    \caption{Number of track intersections per pixel reconstructed based on a triplet reconstructed in the first three telescope planes and matched to a hit in the first telescope plane behind the EoS card.}
    \label{fig:map_tracks}
\end{figure}

A first comparison of the material description was conducted by extrapolating tracks found in the first three telescope planes onto the first telescope sensor behind the EoS card. The closest hit within a window of \unit[$\pm0.8$]{mm} from the extrapolated track intersection was matched. The map of track intersections in figure~\ref{fig:map_tracks} shows a smooth beam profile. This map is independent of the applied material description and therefore identical for all investigated cases.

\begin{figure}[htp]
    \centering
    \includegraphics[width=0.8\linewidth]{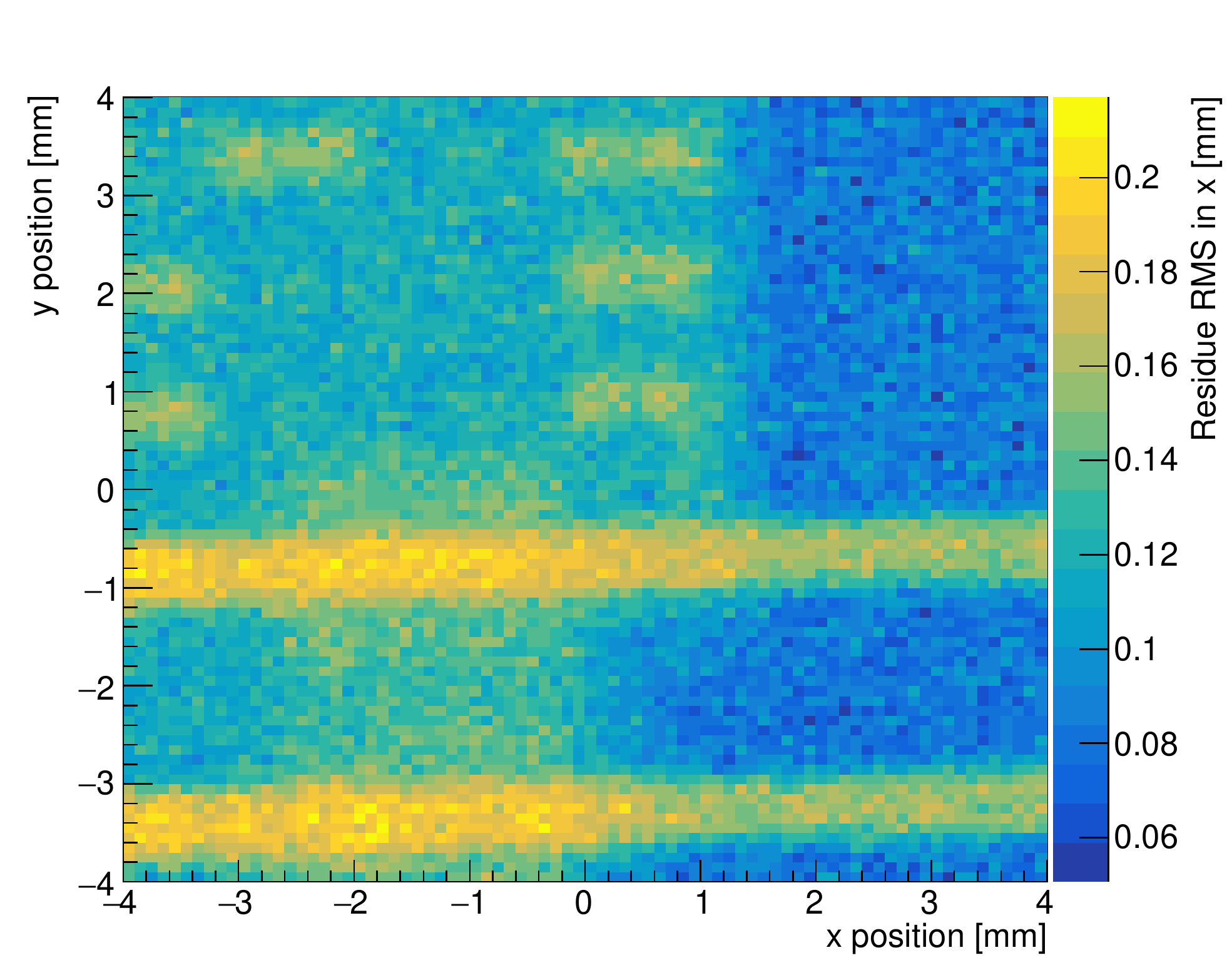}
    \caption{Map of the RMS of residuals per pixel for tracks reconstructed based on a triplet in the first three telescope planes and matched to a hit in the first telescope plane behind the EoS card.}
    \label{fig:map_rms}
\end{figure}

The standard deviation of position residuals $x_{\mathrm{fit}}-x_{\mathrm{hit}}$ has a multiple scattering component that scales with the square root of the traversed material at the EoS card. The non homogeneous material distribution at the EoS card is indeed  visible when mapping the standard deviation of residuals, see figure~\ref{fig:map_rms}.

\begin{figure}[htp]
    \centering
    \includegraphics[width=\linewidth]{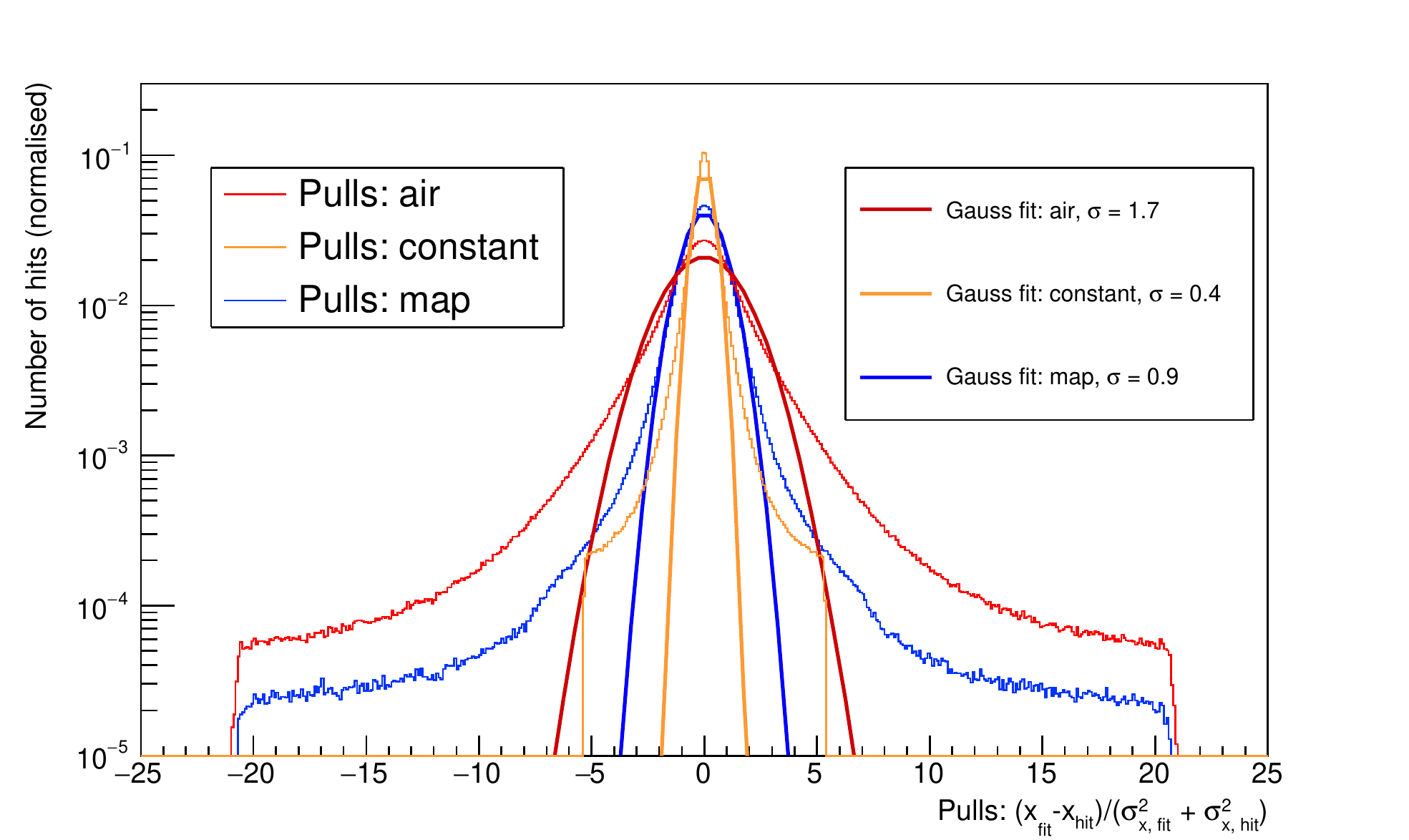}
    \caption{Comparison of pull values for hits in the EoS plane using three different material models for the EoS card. The width of the distributions illustrates that the air model underestimates the scattering material, while the constant distribution overestimates it in areas. The map model provides the most realistic estimate.}
    \label{fig:pulls_comp}
\end{figure}

The improved material description by a detailed $X/X_0$ map leads to an improved estimation of the Kalman filter for the variance of extrapolated track intersections $x_{\mathrm{fit}}$. Figure~\ref{fig:pulls_comp} shows the distribution of standardised residuals (pulls) for all three cases (material descriptions as air, constant and map). 

\begin{figure}[htp]
    \centering
    \includegraphics[width=\linewidth]{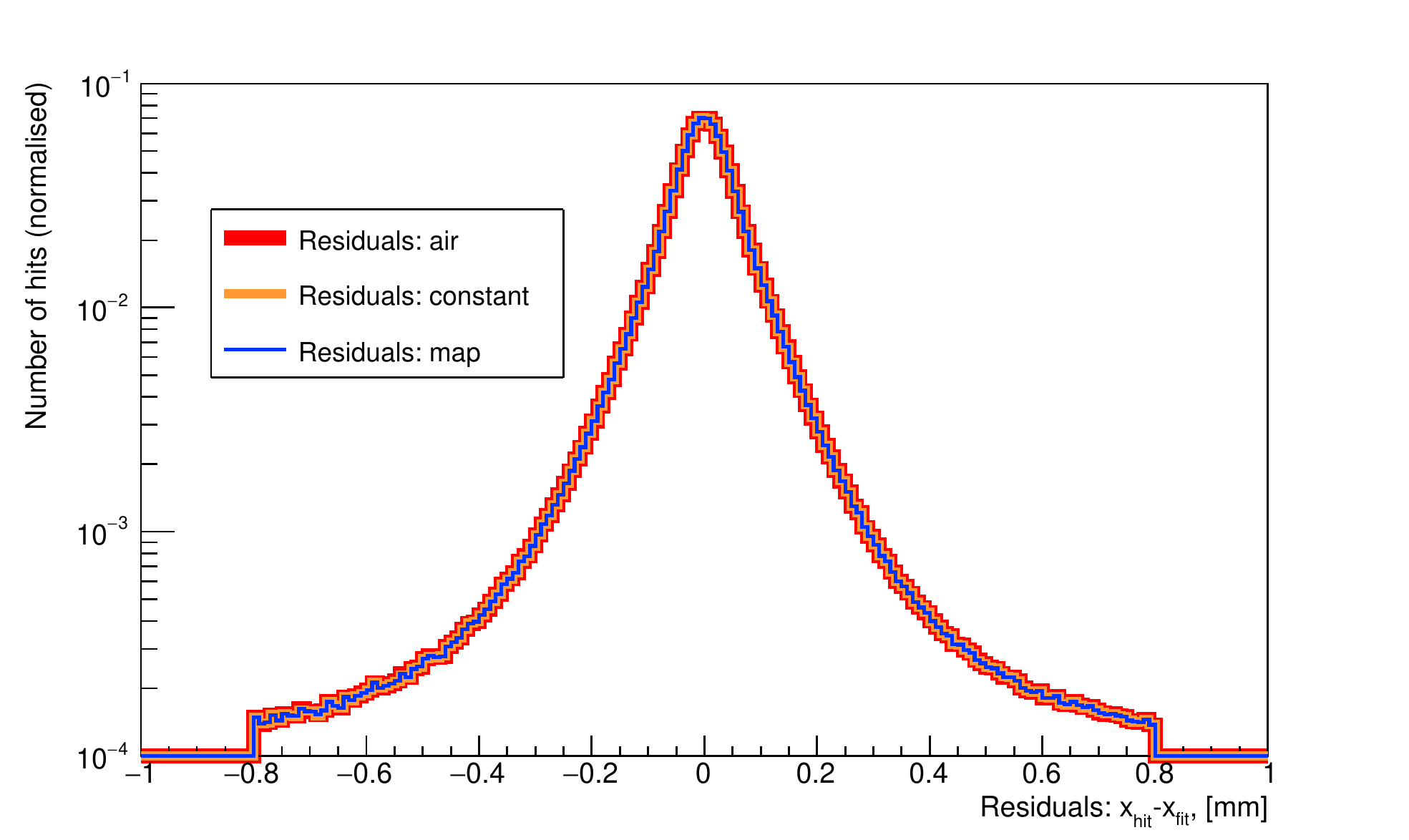}
    \caption{Residual distribution for all hits in a track sample reconstructed from the same data set using different material descriptions. As expected, the residuals are not affected by the material description of the EoS card.}
    \label{fig:res_comp}
\end{figure}

A Gaussian fit to the standardised residual histogram shows that the map description leads to the best fit to a normal distribution and the fitted standard deviation is closest to one, i.e. provides the best estimate for the uncertainty of extrapolated hit positions. In all three cases non Gaussian tails are observed in the standardised pulls. These tails are already present in the residual histogram (see figure ~\ref{fig:res_comp}) and likely  originate from non Gaussian multiple scattering or wrongly matched hits during track finding.  

\begin{figure}[htp]
    \centering
    \includegraphics[width=\linewidth]{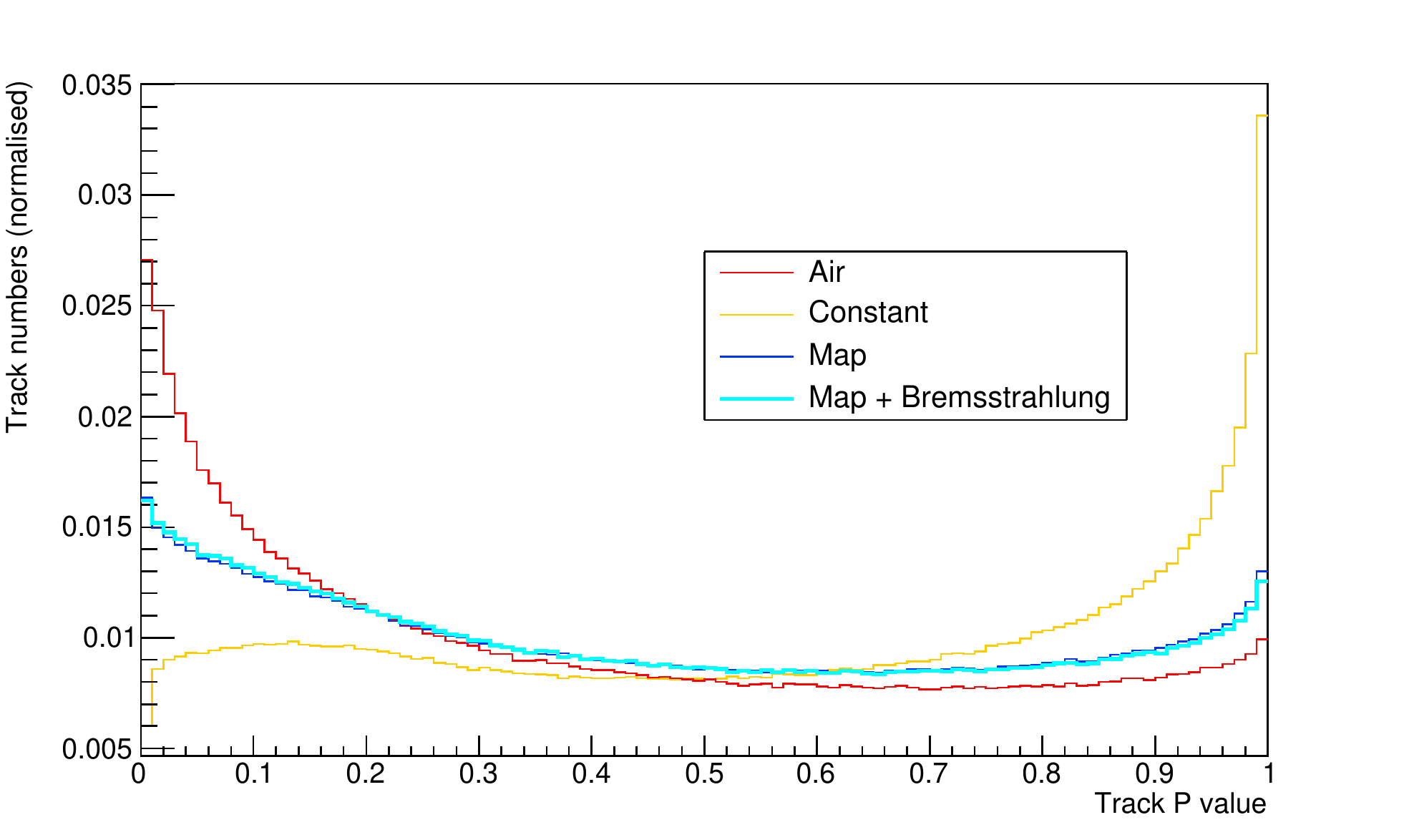}
    \caption{Frequency distributions of the $\chi^2$-probabilities for tracks reconstructed from the same data set using different material descriptions. Due to the different numbers of reconstructed tracks, the distributions were normalised for comparability. As expected, the air sample shows a large fraction of tracks with an underestimated scattering angle, while the flat estimate of the constant distribution, based on the full EoS card with several areas with high material density, overestimates the material for a large fraction of tracks in this subset of data. The Kolmogorov-Smirnov test was applied to quantify the difference between a map description with and without inclusion of a bremsstrahlung correction (see table~\ref{table:Kolmo}).}
    \label{fig:trackPval}
\end{figure}

Only tracks with hits on all six sensors and a total track $\chi^2$-value below 40 where accepted. In addition, new hits were added to the track if the $\chi^2$-value of the predicted position residual was $<10$. It should be noted that the material description at the central EoS card determined if a hit from a downstream sensor was added to the track. As a consequence, the tracking was repeated for all three material descriptions, which led to three different track samples:
\begin{itemize}
    \item air: 2.4 million tracks
    \item constant: 3.5 million tracks
    \item map: 3.4 million tracks
\end{itemize}
A material description as air was found to result in many missed tracks, because large scatterings at the EoS card resulted in high $\chi^2$ values due to the underestimated material at the EoS card. The $\chi^2$ probability distribution of the track fit is shown in figure~\ref{fig:trackPval}.

For a material description as a map, the distribution is closest to a uniform distribution in $[0,1]$ (see table~\ref{table:Kolmo}). While the map description was found to be an improvement, multiple scattering and energy loss by bremsstrahlung are non Gaussian physical processes and provide only an approximation to actual scattering processes, resulting in deviations from a uniform distribution.
\begin{table}[htp]
	\centering
	\begin{tabular}{lc} \toprule
	 	Model & P-value \\ \midrule
		Air & 0.06 \\
		Constant & 0.38 \\
		Map, without bremsstrahlung & 0.60 \\
	    Map, with bremsstrahlung & 0.68 \\
		\bottomrule
	\end{tabular}
	\caption{P-values of the Kolmogorov-Smirnov test comparing the observed distribution of the $\chi^2$-probabilities of the track fit with the ideal flat distribution. As expected, the P-value improves with improved material description.}
	\label{table:Kolmo}
\end{table}

This study confirmed that implementing the reconstructed material distribution for tracking reduces over- or underestimating track uncertainties. It also helps to avoid the loss of tracks due to an underestimate of scattering material. The study confirmed that the explicit description of non homogeneous materials provides an advantage over using averaged material. 

\section{Conclusion and Outlook}

The studies presented here were conducted to test the radiation length imaging algorithm, developed for electron beams, for structures of different complexities. Test structures with a range of material thicknesses were used to expand the existing algorithm for measurements of objects with thicknesses of up to \unit[40]{\%} $X/X_{0}$. The measurements on wedges motivated the implementation of an empirical correction for the energy loss of electrons by bremsstrahlung when passing the device under study. 

The correction was tested further in measurements of an EoS card for the ATLAS ITk as a structure with highly inhomogeneous material distribution. The measurements found a good agreement between the present material, where known, and the reconstructed values.

The measurements investigated the impact of different parameters on the reconstructed material distribution and found reliable working points for parameters such as stepping size between stage positions for adjacent images, minimum number of required tracks per pixel and minimum number of events to be used for analysis per image. The measurements confirmed that the reconstruction of scattering angles for an object consisting of different layers, e.g. the wires surrounded by isolation material and the PCB with embedded copper layers, results in the expected overall material budget. In contrast, areas with highly inhomogeneous material distributions within one reconstructed image pixel showed a systematic shift in the reconstructed material. A single Gaussian distribution was found insufficient to describe more complex material distributions within one imaged pixel, which can be avoided by choosing a pixel size below the feature size of the imaged object.

Future modifications of the analysis code used here are therefore planned to incorporate a double Gaussian fit which does take mixed contributions within the same pixel into account: instead of implementing a single fit to approximate mixed contributions, using a fit function that contains Gaussian contributions with different distributions widths has been found to produce more reliable results for pixels with mixed contributions.

\section*{Acknowledgements}

The work presented here was partially funded by the Canada Foundation for Innovation (CFI) and the Natural Sciences and Engineering Research Council (NSERC) of Canada as well as the Alexander von Humboldt Foundation. We acknowledge the financial support by the Federal Ministry of Education and Research of Germany.

The measurements leading to these results have been performed at the Test Beam Facility at DESY Hamburg (Germany), a member of the Helmholtz Association (HGF). The authors would like to thank the DESY testbeam coordinators and beam telescope support, especially Dr. Jan Dreyling-Eschweiler.

\bibliographystyle{unsrt}
\bibliography{bibliography.bib}

\begin{thebibliography}{10}

\bibitem{Cooper}
William~E. Cooper.
\newblock {Low-Mass Materials and Vertex Detector Systems}.
\newblock {\em PoS}, Vertex2013:036, 2013.

\bibitem{Aaboud:2017pjd}
{ATLAS~Collaboration}.
\newblock {Study of the material of the ATLAS Inner Detector for Run 2 of the
  LHC}.
\newblock {\em JINST}, 12(12):P12009, 2017.

\bibitem{Sirunyan_2018}
A.M.~Sirunyan et. al.
\newblock {Precision measurement of the structure of the {CMS} inner tracking
  system using nuclear interactions}.
\newblock {\em {Journal of Instrumentation}}, 13(10):P10034--P10034, oct 2018.

\bibitem{demarteau}
M.~Demarteau et. al.
\newblock {Planning the Future of U.S. Particle Physics (Snowmass 2013):
  Chapter 8: Instrumentation Frontier}, 2014.

\bibitem{Stolzenberg2019}
U.~Stolzenberg.
\newblock {\em {Radiation length measurements with high-resolution
  telescopes}}.
\newblock PhD thesis, Universit{\"a}t G{\"o}ttingen, 2019.
\newblock {II.Physik-UniGö-Diss-2019/06}.

\bibitem{vci_2017_x0}
U.~Stolzenberg, A.~Frey, B.~Schwenker, P.~Wieduwilt, C.~Marinas, and
  F.~Lütticke.
\newblock {Radiation length imaging with high-resolution telescopes}.
\newblock {\em {Nuclear Instruments and Methods in Physics Research Section A:
  Accelerators, Spectrometers, Detectors and Associated Equipment}}, 845:173 --
  176, 2017.
\newblock Proceedings of the Vienna Conference on Instrumentation 2016.

\bibitem{EOS}
C.~{Wanotayaroj}, H.~{Ceslik}, H.~{Colbow}, S.~D. {Cornell}, P.~{Goettlicher},
  I.~M. {Gregor}, A.~{Melnik}, M.~{Stanitzki}, and J.~{Wolff}.
\newblock {The End-of-Substructure (EoS) Card for the Strip Tracker Upgrade of
  the ATLAS experiment}.
\newblock In {\em {2018 IEEE Nuclear Science Symposium and Medical Imaging
  Conference Proceedings (NSS/MIC)}}, pages 1--3, 2018.

\bibitem{ITk}
{ATLAS~Collaboration}.
\newblock {Technical Design Report for the ATLAS Inner Tracker Strip Detector}.
\newblock Technical Report CERN-LHCC-2017-005. ATLAS-TDR-025, CERN, Geneva, Apr
  2017.

\bibitem{ATLAS}
{ATLAS~Collaboration}.
\newblock {The ATLAS Experiment at the CERN Large Hadron Collider}.
\newblock {\em Journal of Instrumentation}, 3(08):S08003, 2008.

\bibitem{pdg_1998}
C.~Caso et~al.
\newblock {Review of Particle Physics}.
\newblock {\em Eur.Phys.J.}, C3, 1998.

\bibitem{Moliere_multiple_scattering}
G.~Moli{\`e}re.
\newblock {Theorie der Streuung schneller geladener Teilchen II. Mehrfach- und
  Vielfachstreuung}.
\newblock {\em Zeitschrift Naturforschung Teil A}, 3:78--97, 1947.

\bibitem{Highland_1975}
V.~Highland.
\newblock {Some practical remarks on multiple scattering}.
\newblock {\em {Nucl. Instr. and Meth.}}, 129(2):497 -- 499, 1975.

\bibitem{pdg_2018}
M.~Tanabashi, K.~Hagiwara, and Hikasa.
\newblock {Review of Particle Physics}.
\newblock {\em Phys.Rev.D}, 98, 2018.

\bibitem{Lynch_1991}
G.~R. Lynch and O.~I. Dahl.
\newblock {Approximations to multiple Coulomb scattering}.
\newblock {\em {Nucl. Instr. Meth.}}, B58:6 -- 10, 1991.

\bibitem{DESYII}
R.~Diener, J.~Dreyling-Eschweiler, H.~Ehrlichmann, I.M. Gregor, U.~Kötz,
  U.~Krämer, N.~Meyners, N.~Potylitsina-Kube, A.~Schütz, P.~Schütze, and
  M.~Stanitzki.
\newblock {The DESY II test beam facility}.
\newblock {\em {Nuclear Instruments and Methods in Physics Research Section A:
  Accelerators, Spectrometers, Detectors and Associated Equipment}}, 922:265 --
  286, 2019.

\bibitem{tbsw}
Testbeam~Software Developers, 2016.
\newblock Test Beam Software Framework (TBSW),
  \url{https://bitbucket.org/testbeam/tbsw}.

\bibitem{Rubinskiy2012923}
I.~Rubinskiy.
\newblock {An EUDET/AIDA Pixel Beam Telescope for Detector Development}.
\newblock volume~37, pages 923 -- 931, 2012.
\newblock {Proceedings of the 2nd International Conference on Technology and
  Instrumentation in Particle Physics (TIPP 2011) }.

\bibitem{Schwenker14}
B.~Schwenker.
\newblock {\em {Development and validation of a model for the response of the
  Belle II vertex detector}}.
\newblock PhD thesis, Universit{\"a}t G{\"o}ttingen, 2014.
\newblock {II.Physik-UniGö-Diss-2014/05, CERN-THESIS-2014-165}.

\bibitem{BetheHeitler}
Heitler~W. Bethe~H. and Dirac Paul~Adrien Maurice.
\newblock {On the stopping of fast particles and on the creation of positive
  electrons}.
\newblock {\em Proc. R. Soc. Lond. A}, 146:83–112, 1934.

\bibitem{GBTX}
P~Leitao, S~Feger, et~al.
\newblock {Test bench development for the radiation hard {GBTX} {ASIC}}.
\newblock {\em {Journal of Instrumentation}}, 10(01):{C01038--C01038}, {Jan}
  2015.

\bibitem{GBT-SCA}
{A. Caratelli and S. Bonacini and K. Kloukinas and A. Marchioro and P. Moreira
  and R. De Oliveira and C. Paillard}.
\newblock {The {GBT}-{SCA}, a radiation tolerant {ASIC} for detector control
  and monitoring applications in {HEP} experiments}.
\newblock {\em {Journal of Instrumentation}}, 10(03), mar 2015.

\bibitem{DATURA}
{Jansen, H. and Spannagel, S. and Behr, J. and others}.
\newblock {Performance of the EUDET-type beam telescopes}.
\newblock {\em {EPJ Techniques and Instrumentation}}, 3(7), 2016.

\bibitem{MIMOSA}
J.~Baudot et~al.
\newblock {First test results of MIMOSA-26, a fast CMOS sensor with integrated
  zero suppression and digitized output}.
\newblock In {\em 2009 IEEE Nuclear Science Symposium Conference Record
  (NSS/MIC)}, pages 1169--1173, 10 2009.

\end{thebibliography}

\end{document}